\documentclass[pdflatex,sn-nature]{sn-jnl}

\usepackage{graphicx}%
\usepackage{multirow}%
\usepackage{amsmath,amssymb,amsfonts}%
\usepackage{amsthm}%
\usepackage{mathrsfs}%
\usepackage[title]{appendix}%
\usepackage{xcolor}%
\usepackage{textcomp}%
\usepackage{manyfoot}%
\usepackage{booktabs}\let\cline\cmidrule%
\usepackage{listings}%
\usepackage{dcolumn}
\usepackage{float}
\usepackage{longtable}
\usepackage[english]{babel}
\usepackage[autostyle, english = american,maxlevel=4]{csquotes}
\MakeOuterQuote{"}

\begin{document}

\title[]{A survey of generative AI adoption and perceived productivity among scientists who program}


\author*[1]{\fnm{Gabrielle} \sur{O'Brien}}\email{elleobri@umich.edu}

\author[1]{\fnm{Alexis} \sur{Parker}}\email{paralexi@umich.edu}

\author[2]{\fnm{Nasir U.} \sur{Eisty}}\email{neisty@utk.edu}

\author[3]{\fnm{Jeffrey} \sur{Carver}}\email{carver@cs.ua.edu}

\affil*[1]{\orgdiv{School of Information}, \orgname{University of Michigan}, \orgaddress{\street{2200 Hayward St}, \city{Ann Arbor}, \postcode{48109}, \state{Michigan}, \country{United States}}}

\affil[2]{\orgdiv{EECS}, \orgname{University of Tennessee}, \orgaddress{\street{1203 W. Cumberland Ave}, \city{Knoxville}, \postcode{37996}, \state{Tennessee}, \country{United States}}}

\affil[3]{\orgdiv{Computer Science}, \orgname{University of Alabama}, \orgaddress{\street{Box 870290}, \city{Tuscaloosa}, \postcode{35487}, \state{AL}, \country{USA}}}


\abstract{Programming is essential to modern scientific research, yet most scientists report inadequate training for the software development their work demands. Generative AI tools capable of code generation may support scientific programmers, but user studies indicate risks of over-reliance, particularly among inexperienced users. We surveyed 868 scientists who program, examining adoption patterns, tool preferences, and factors associated with perceived productivity. Adoption is highest among students and less experienced programmers, with variation across fields. Scientific programmers overwhelmingly prefer general-purpose conversational interfaces like ChatGPT over developer-specific tools. Both inexperience and limited use of development practices (like testing, code review, and version control) are associated with greater perceived productivity---but these factors interact, suggesting formal practices may partially compensate for inexperience. The strongest predictor of perceived productivity is the number of lines of generated code typically accepted at once. These findings suggest scientific programmers using generative AI may gauge productivity by code generation rather than validation.}


\maketitle
\section{Introduction}\label{intro}

The recent emergence of generative AI (genAI)-based tools capable of code generation
marks a potentially transformative inflection point in how programs are written and
by whom \cite{Cui2025TheDevelopers, Peng2023TheCopilot, DellAcqua2023NavigatingQuality}. In modern scientific research, programming underpins knowledge work well
beyond traditionally computational domains \cite{Hettrick2014ItsResearchers,
Wilson2016SoftwareLearned}. Because the demand for programming support frequently
exceeds scientists' training and resources \cite{Carver2022AStates,
Nangia2017SurveyingResearch}, advances in genAI for software development stand to
affect the day-to-day work of scientific research broadly.

The same factors that make scientific programmers potential beneficiaries of these
tools also present avenues for risk. Observational studies of programmers working
with genAI tools indicate that over-reliance can occur, in which users uncritically
accept generated code without verifying or understanding
it~\cite{Prather2024TheProgrammers, MoradiDakhel2023GitHubLiability,
Mozannar2024ReadingProgramming, Vaithilingam2022ExpectationModels,
Barke2023GroundedModels}. Recently, researchers at Anthropic, a major LLM provider, concluded from a randomized experiment of programmers that the use of a conversational assistant ``impairs conceptual understanding, code reading, and
debugging abilities, without delivering significant efficiency gains on
average''~\cite{Shen2026HowFormation}. Because most scientists report receiving little programming
education~\cite{Carver2022AStates, Nangia2017SurveyingResearch,
Hettrick2014ItsResearchers}, those who adopt genAI tools to compensate could
paradoxically find themselves facing new threats to learning.

A further tension is that the work genAI makes most critical---evaluating the
soundness and correctness of generated software~\cite{Yetistiren2023EvaluatingChatGPT,
MoradiDakhel2023GitHubLiability}---is one that scientific software development has
historically struggled with~\cite{Sanders2008DealingDevelopment,
Kelly2015ScientificSoftware, Vogel2019ChallengesScience}. Seemingly minor
mistakes in scientific code can have radical consequences for how results are
understood and interpreted (as documented cases of code-related
retractions illustrate~\cite{2016Retraction:2016, Karraker2015AuthorsRetraction,
Mandhane2024Notice202317711:1226-1228.}), with the potential to mislead research
directions or damage public trust. Yet development practices that support code
quality---like testing or review---are used highly inconsistently in research
settings~\cite{Carver2022AStates, Wilson2017GoodComputing,
Nguyen-Hoan2010ADevelopment, Prabhu2011AScience} (though we note that scientists
with high software standards may also engage in flexible, exploration-oriented
programming as a core research activity~\cite{Sutherland-Keller2025ResearchCosmology}).
If code is increasingly generated without close oversight, and without pre-existing
infrastructure for managing its volume and correctness, the historically uneasy
relationship between scientific software and validation work may be further strained.

Despite the potential for widespread effects on how science is practiced, empirical
data on genAI adoption among scientists for programming remains sparse. This study
addresses three aspects of adoption using survey methods. First, how common is the
use of genAI tools for research-related programming, and which tools are most widely
used? Understanding adoption patterns across career stages, fields, and experience
levels is a prerequisite to gauging potential impacts. Second, what factors are
associated with perceived productivity with genAI tools? Prior work in general
developer populations suggests that less experienced developers perceive the greatest
productivity gains~\cite{Ziegler2022ProductivityCompletion}, but whether this holds
for scientific programmers is untested. We also explore a hypothesis motivated by
the potential for automation bias~\cite{Goddard2012AutomationMitigators} and
over-reliance~\cite{Prather2024TheProgrammers, OBrien2025HowProgram}: that
scientists who lack existing processes for code oversight---like testing or
review---will be especially likely to rate themselves as highly productive with
genAI tools. Third, why do some scientists reject genAI tools for programming? Our
cross-sectional survey design does not allow definitive claims about how perceived
productivity relates to software quality or correctness, but it allows us to
characterize adoption patterns and their correlates in a population where such
questions have not previously been examined.

\subsection{Related work}

While studies directly targeting scientific programmers' use of genAI tools are
limited, a broader literature addresses how programmers generally work with genAI
tools, the state of development practices in scientific settings, and the adoption
of genAI in research more broadly. We briefly review each in turn.

\subsection{Programming with genAI tools}

Since genAI tools like ChatGPT and GitHub Copilot became widely available, numerous
studies have characterized how they are changing software development, mostly in
non-scientific contexts. Some suggest that perceptions of increased developer
productivity correspond to measurable increases in programming
activity~\cite{Weber2024SignificantModels, Kumar2025IntuitionProductivity}, though
this is inconsistent---other studies have found no meaningful productivity increases
or even observed reductions~\cite{BeckerMeasuringProductivity,
Stray2025DeveloperStudy}. Increased developer velocity, even when observed, may be
accompanied by subsequent declines in code quality that complicate
maintenance~\cite{He2025SpeedDevelopment}.

A large study of GitHub Copilot users cross-referencing survey data with activity
logs found that less experienced developers attributed the greatest productivity gains
to the tool~\cite{Ziegler2022ProductivityCompletion}. Perceived productivity was also
strongly related to how often a user accepted code suggestions. Several mechanisms
could explain this: novice programmers may genuinely benefit most from assistance,
or they may be insufficiently critical of suggestions~\cite{Prather2024TheProgrammers}.
Prather et al.~\cite{Prather2024TheProgrammers} described the ``illusion of
competence'' among novice programmers using genAI tools, finding they greatly
overestimated both their understanding of generated code and their own programming
ability. This has antecedents in earlier eras of programming tools: spreadsheet
programmers have similarly been observed to overestimate the correctness of
automatically produced values~\cite{Panko2008TwoDevelopment, KoTheEngineering}.

Users who frequently accept generated code may be making conscious choices to delay
verification~\cite{Mozannar2024ReadingProgramming}. However, some studies suggest
that inexperienced programmers in particular may rely on superficial checking or
believe verification is unnecessary~\cite{Fawzy2025VibeReview, OBrien2025HowProgram,
Nguyen2024HowOther, Prather2023ItsProgrammers}. Research on genAI use in other
knowledge work contexts similarly finds that many users avoid evaluating outputs,
especially when confidence in the tool is high~\cite{Lee2025TheWorkers}. We seek to
understand what factors are associated with perceived productivity among scientific
programmers, while acknowledging that our cross-sectional design does not allow
conclusions about how these perceptions relate to software quality or correctness.

\subsection{Scientific programming practices}

Although there have been considerable strides in professionalizing scientific
software~\cite{Teal2015DataResearchers, Wilson2016SoftwareLearned,
Wilson2017GoodComputing}, substantial variability remains in where infrastructure
and expertise are available to support the practices that serve as common guardrails
on code quality in industry settings~\cite{Carver2022AStates, Wilson2017GoodComputing,
Nguyen-Hoan2010ADevelopment, Prabhu2011AScience}. Practices like code
testing~\cite{Eisty2025TestingTools, Kanewala2014TestingReview}, code
review~\cite{ArifulIslamMalik2025PeerPerspective}, and version
control~\cite{JayNotThem} are used inconsistently in research settings, if at all.
This is especially challenging given that some scientific software lacks a ground
truth against which outputs can be verified~\cite{Vogel2019ChallengesScience}.

We take as a premise that development practices scaffold opportunities for deeper
scrutiny of code, and hypothesize that their use will moderate perceptions of
productivity with genAI. This does not, however, exclude the possibility that
scientists engage in verification work that does not fit neatly into the standard
toolkit of development practices. Observations of experienced computational
scientists have shown them treating code with considerable skepticism without
practices that would conventionally be classified as testing~\cite{Kelly2015ScientificSoftware,
Kelly2011ScientificDimensions}, and some scientists deliberately delay validation
during exploratory stages of programming~\cite{Sutherland-Keller2025ResearchCosmology}.
Our survey provides an initial characterization of development practice adoption in
our sample and its relationship to perceived productivity with genAI tools, while
acknowledging that this characterization is not exhaustive.

\subsection{Adoption of genAI tools in science research}

Empirical data on genAI adoption among scientists for programming is currently limited. In a 2023 \textit{Nature} survey, over 40\% of researchers agreed genAI would make programming faster~\cite{VanNoorden2023AIThink}, but actual use was not assessed. A more recent nationally representative survey of academic scientists found that 65\% used AI tools in their research at least some of the time, but did not address programming specifically~\cite{Arroyo-Machado2025GenerativeIntentions}; junior scientists and trainees, who may do much of the programming work, were also not well-represented. A 2025 survey at one large institute reported 43.2\% of
respondents used AI to ``help write code'', again without examining programming in
depth~\cite{Chugunova2025WhoGermany}, and found a substantial gender gap in
adoption, with women more likely to report being unfamiliar with available tools.

A 2024 single-institution study recruited scientists who program with genAI
tools~\cite{OBrien2025HowProgram}, collecting information on tool choice, research
area, and programming experience, but with limited questions about practices and
productivity. Most respondents in that sample used general-purpose tools like
ChatGPT rather than developer-specific tools like GitHub Copilot, consistent with
the broad appeal of conversational interfaces to relatively inexperienced
programmers~\cite{Fawzy2025VibeReview, Treude2025HowEngineering}. There has also been at least one study to date targeting attitudes towards genAI tools in the research software engineering community, suggesting a mixture of caution and curiosity \cite{VanTuylStateConvening}. 

Gaps remain in understanding how common genAI adoption is for
scientific programming across a range of career stages and fields, which tools are preferred, and what factors are associated with adoption and perceived productivity. Our survey is designed to address these
questions in a broad sample of scientists who program.

\section{Methods}\label{section:methods}
We conducted a survey study of development practices and generative AI tool adoption patterns among research scientists who program as part of their scientific work.  We sought broad recruitment across fields and career stages using mailing lists in our professional networks. The survey was designed to target three aspects of adoption we hoped to understand: 

\begin{itemize}
    \item \textbf{Who uses genAI tools most in their research-related programming, and which tools do they prefer?} We collected participants' self-reported usage frequency of genAI tooling, their choice of tools, and demographic information such as research area and years of programming experience.
    \item \textbf{How do users' programming experience and software development practices relate to perceived productivity with genAI tools?} To test our hypotheses about predictors of perceived productivity, we adapted existing survey instruments addressing programmers' perceived productivity with genAI assistance and use of development practices.  
    \item \textbf{Why do some scientists reject genAI tools for programming?} Among participants who indicated they do not regularly program with any genAI tools, we solicited open-ended responses about their reasons for non-adoption. 
\end{itemize}

Prior to dissemination, the survey and study plan were reviewed and approved by the University of Michigan Institutional Review Board. All methods were carried out in accordance with relevant guidelines and regulations, and informed consent was obtained from all subjects.

The full survey is available as supplementary material.

\subsection{Survey design}\label{section:survey-design}
The survey instrument consisted of seven main sections administered through Qualtrics, with an estimated completion time of 10-15 minutes. All responses were anonymous, meaning IP addresses were not collected. The complete survey is provided in the supplemental material. All participants were required to provide consent to participate and confirm their eligibility (18 years of age or older) before beginning. Core survey sections were as follows:

\paragraph{Demographics}
Participants were asked a series of demographic questions about the organization where they work, the country in which they work, and their current job title. Participants were asked to select the best-matching research area from the National Science Foundation's Codes for Classification of Research (e.g., "Life sciences", "Engineering", "Physical sciences", "Education").

\paragraph{Programming background} 
Participants were asked about their years of programming experience, the languages they use, and how often they program in their research. As this study targets scientists who do research-related programming, participants who indicated they never programmed in any language as part of their scientific research were directed to an exit page and screened out of the survey. 

\paragraph{Programming practices}\label{section:methods-practices}
All other participants were then directed to a section on their typical coding practices, which asked about their familiarity and typical usage of a set of software development practices (version control, code testing, code review, and continuous integration) based on a previous survey of scientific programmers \cite{Carver2022AStates}.  This section also asked about code publishing and reuse, both inside and outside the participant's research group.

\paragraph{GenAI Tool Experience}
The following section asked participants about their experiences with GenAI programming tools, including which tools they had experimented with and how often they use any tools in their research-related programming. Participants who indicated that they had not tried any tools or had tried but given up were directed to an open-ended question about why they think they do not use these tools, followed by an exit screen. 

\paragraph{Perceived productivity scale}\label{section:methods-space}
All other participants were directed to the SPACE perceived productivity questionnaire (adapted from a study of perceived productivity in GitHub Copilot users \cite{Ziegler2022ProductivityCompletion}). Participants were asked to rate their agreement with nine statements about how their primary genAI code tool had affected their programming on a 5-point Likert scale. Items are mapped to five dimensions from a theoretical model of developer productivity, satisfaction and well-being, performance, activity, communication and collaboration, efficiency and flow \cite{Forsgren2021TheProductivity}. The aspects of productivity addressed in the questionnaire are reported in Table ~\ref{tab:space_items}. 

One change we made from the original implementation concerned the wording of statements about tool choice, as the original instrument was designed for questions about GitHub Copilot. Because participants in our study used a variety of genAI tools in their research programming, we introduced the productivity section of the survey with framing, \textit{``Think about how you use your primary genAI tool in research-related programming.''} This framing was scaffolded by preceding questions about their primary tool use in the previous section (\textit{``You indicated you have tried the following genAI tools for programming support. Which one do you use MOST in your research-related programming?''}, \textit{``How do you typically access this tool?''}, \textit{``How many lines of generated code suggestions do you typically accept at one time when working with this tool?''}). In the SPACE questionnaire items, we replaced language from the original wording to be applicable to whichever primary tool the respondent used (for example, the statement ``I learn from the suggestions GitHub Copilot shows me'' became ``I learn from the responses this tool gives me''). 

We also added a statement addressing the "A" dimension of the SPACE model, activity. In the GitHub Copilot study, user telemetry was available via logging, and this was considered a superior estimator of activity than self-reporting. Because we do not have access to telemetry for the tools used by participants, we created an item based on the SPACE theoretical model addressing activity: "I produce more lines of code with this tool than I would have without it." Additionally, we added an item to dimension C (communication and collaboration) about tools that facilitate understanding other people's code, as interviews with scientific programmers who use genAI tools have surfaced this as a common use case \cite{OBrien2025HowProgram}. To avoid an overly long instrument, and noting the correlation structure of responses to items from the GitHub Copilot study, we reduced the number of items on the S and E (satisfaction and well-being, and efficiency and flow) items to two each. The choice of items was based on which seemed most relevant to scientific programmers and which items showed the strongest clustering within the intended factor. 

\paragraph{Example use case}
After the productivity scale was administered, participants were asked to provide an example use case and their strategy for evaluating the output in a series of open-ended questions (adapted from \cite{Lee2025TheWorkers}). Note that this data is not presented here for manuscript length reasons.

\bigskip
Regardless of how participants answered questions about their GenAI tool experience, all were invited to share any other thoughts on the survey topic in an open-ended text box, and to optionally provide contact information to participate in follow-up research or be entered in the prize drawing.

Because some information collected in our survey could be sensitive (i.e., usage, or not, of software development practices, or sharing opinions about genAI tools), we took care to ensure the anonymity of responses and give participants the option to skip any questions. This means that the number of responses to any given question may be less than the total number of participants. Additionally, we did not collect IP addresses attached to any responses. 

\subsection{Survey recruitment}\label{section:methods-recruitment}
Invitations to participate in the survey were spread through several mailing lists and online communities for scientific programming in the author's professional networks. This included Slack groups for the United States Research Software Engineering  (US-RSE) professional association, the pyOpenSci community, and an internal research programming group at the University of Michigan. We also recruited via the monthly email newsletters of the Alfred P. Sloan Foundation's Digital Technology Program and the US-RSE organization, as well as the Learning Engineering Google group. Finally, invitations were sent via a targeted mailing list at the University of Michigan to employees with “research” in their job title (18,851 addresses). Because the survey link was shareable, recipients may also have shared it within their networks. As such, it is not possible to definitively calculate a response rate. Survey participation was incentivized by the option to enter a drawing for one of 5 cash prizes (\$100 each). 

\subsubsection{Inclusion criteria}\label{section:methods-inclusion}
In total, 1272 responses were collected between July 10 and August 25, 2025. Before analysis, we removed any responses that did not complete the survey (283) or answer affirmatively to three informed consent questions (8). As part of the survey’s screening questions, participants were asked how often they program as part of their scientific research; 102 respondents indicated they never program in this capacity and were screened out of the survey. One additional quality control measure was added early in the data collection process: within the first 40 responses, an experimenter noticed that one participant working in the private sector described an off-topic (non-scientific) programming use case. This response was filtered out for quality control, and a screening question was added: “Does your research group publish peer-reviewed works of research, like journal articles or conference proceedings?” 9 responses that answered “no” to this question were screened out. This left 868 responses to be analyzed.

\subsection{Quantitative data handling}\label{section:data-handling}
For reproducibility, R scripts for processing data from the survey are provided in the accompanying GitHub repository. Responses not meeting the inclusion criteria above were dropped programmatically, and any skipped questions were coded as “NA” values. Questions asking participants to give ratings on a scale (i.e., how often they program, use a development practice, or use a genAI tool in programming) were transformed into ordered factors for analysis. Years of programming and research experience variables were $\log_{10}$ transformed, with an offset of 1 to prevent undefined values. Finally, we used a standardized dictionary to recode certain job titles that participants had written in, rather than selecting one of the survey choices: several participants had written in some variant of “graduate research assistant” or “PhD student”, which we coded to the “student research assistant” option. The dictionary is provided in the code repository.

All data processing, visualizations, and analyses were conducted in R \cite{RCoreTeam2025R:Computing}. Scripts are provided in the accompanying GitHub repository. Because responses may be incomplete for any particular question as we did not require responses to every question, we used complete pairwise observations for calculating any correlations presented in this manuscript rather than dropping participants with missing data. All correlations are Pearson except where otherwise indicated, as for example in polychoric correlations between Likert scale items. Linear model tables are formatted with the \verb|stargazer| \cite{Hlavac2022Stargazer:Tables} library and correlation matrices are visualized with \verb|corrplot| \cite{Wei2024Corrplot:Matrix}. 

\subsection{Qualitative analysis of non-adoption reasons}\label{section:qualitative-methods}
We analyzed responses to the free-text item asking participants to describe their reasons for not adopting generative AI tools. The survey question read: \textit{“You indicated that you haven't tried any genAI tools, or don't regularly use them, for your research-related programming. Why do you think this is?”} This prompt generated 210 written responses, and each response could include one or more themes. Our analysis focused only on this item.

AP developed an initial open codebook by identifying patterns across the responses and applying those codes to all entries. GO separately reviewed the same responses using that codebook, and then both annotators met to compare our interpretations. At this stage, we began organizing and refining the codes by clarifying meanings, combining similar ideas, and separating those that described different kinds of reasoning. For example, one of the labels created during open coding, “Not needed,” referred generally to participants who saw no reason to use AI tools. On closer review, we found two distinct types of reasoning associated with this theme: some participants felt their programming work was not sufficiently demanding to require such tools, while others found AI tools redundant with the resources they already used (like Stack Overflow). 

GO then created an axial codebook reflecting the refined codes after discussion, and re-labeled all conversations using the axial codes. AP reviewed the axial codes assigned by GO, and any disagreements were resolved through discussion such that perfect agreement was reached. The axial codebook is presented as Table~\ref{tab:axial_codebook} with example responses and frequencies for each theme. All qualitative results presented in the paper are derived from this codebook.

\section{Results}\label{results}

To contextualize our findings, we begin with reporting the characteristics of respondents.

\subsection{Survey response demographics}
Our survey of scientific programmers yielded 868 respondents that met the inclusion criteria, which required programming at least sometimes as part of their research work (Methods~\ref{section:methods-inclusion}). We emphasize that this is a non-random, convenience sample.

The sample skewed heavily toward early-career researchers: graduate student research assistants were the most common respondents (331), followed by research staff (225), faculty (152), and post-doctoral researchers (126). Only 21 respondents identified as research software engineers. This was somewhat surprising, as the survey was advertised in a monthly newsletter and Slack workspace for the United States Research Software Engineer Association. One explanation may be that community members who identify with research software engineering hold a variety of job titles, such as graduate student researcher or research staff, and may prioritize those identities when asked about their position. Overall, however, the sample appears to capture more "scientists who program" than professional scientific programmers. 
 
Programming experience ranged from $<1$ to 55 years (median = 7, $\mu$ = 9.50, $\sigma$ = 9.20; Figure~\ref{fig:years_program_exp}), with the distribution skewing toward less experienced programmers consistent with a high proportion of students. As expected, programming experience was moderately correlated with research experience ($r = 0.54$, $p < 2.2e-16$ for $\log_{10}$-transformed years; note this transformation is applied because of the long-tail of the experience distribution visible in Figure~\ref{fig:years_program_exp}). Python was the most commonly used programming language, followed by R and MATLAB (Table \ref{tab:programming_languages}). There were more men than women in the sample. Of those who provided gender information, 441 identified as men, 384 as women, and 24 as non-binary or gender diverse; 19 respondents declined to answer. 

Life Sciences was the most represented field (264 respondents), followed by Engineering (164), Social Sciences (104), Physical Sciences (89), and Computer and Information Services (57). Respondents were asked to select the best description of their research area based on NSF Codes for Classifications of Research; full counts for all research classifications appear in Table~\ref{tab:research_area}.

Nearly all respondents (809) worked at higher education institutions, with smaller numbers at national laboratories (21), private sector organizations (13), or other settings (23). Geographic representation was heavily skewed toward the United States (843 respondents), with only 25 responses from 13 other countries. 

To summarize, our sample largely reflects academic researchers at universities in the United States, with an especially high representation of early-career scientists (many of whom are graduate students). The majority are situated in disciplines that are not primarily focused on computing, and few respondents identify themselves as research software engineers. We infer from these patterns that this sample largely comprises scientists who code as part of their research, rather than scientists whose primary occupation is developing research software. 

\begin{table}
\caption{Survey respondents' research areas. Respondents were asked to select the best-matching category.}
\label{tab:research_area}
\begin{tabular}{lr} 
Research Area & Count\\
\midrule
Life sciences & 264\\
Engineering & 164\\
Social sciences & 104\\
Physical sciences & 89\\
Computer and information services & 57\\
Other & 52\\
Mathematics and statistics & 40\\
Psychology & 40\\
Geosciences, atmospheric sciences \& ocean sciences & 33\\
Education & 11\\
Business management \& administration & 7\\
Humanities & 4\\
Law & 1\\
Social work & 1\\
\bottomrule
\end{tabular}
\end{table}

\subsection{Who uses genAI tools for programming most?}

To answer the first of our guiding research questions, we asked participants how often they use genAI tools in their research-related programming. We present the overall findings and then examine factors related to relative differences in usage frequency. 

\begin{figure}[h]
\centering
\includegraphics[width=0.9\textwidth]{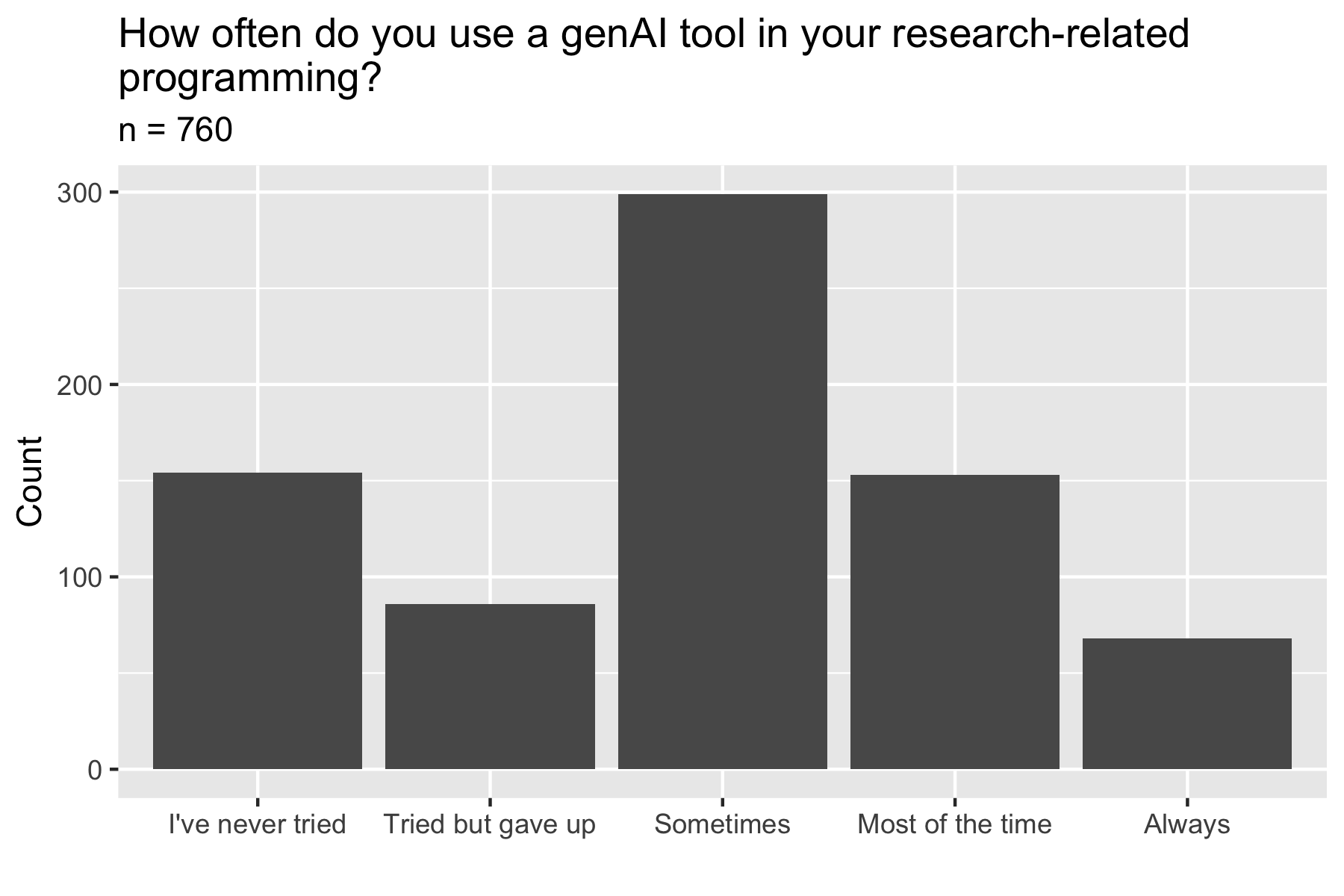}
\caption{Distribution of 760 responses to survey question, \textit{How often do you use a genAI tool in your research-related programming?} }\label{fig:adoption-barchart}
\end{figure}

Overall usage frequency is shown in Figure~\ref{fig:adoption-barchart} (n=760; note that respondents could opt not to answer any survey questions, so we report the total number of responses for each item discussed). The modal response was "Sometimes", and usage patterns split roughly evenly on either side of the distribution around the mode: a similar fraction of respondents programmed "most of the time" or "always" with genAI, as had never tried or "tried but gave up". This heterogeneity suggests that genAI adoption among scientists for programming remains far from universal, although the majority have at least experimented.

\begin{figure}[h]
\centering
\includegraphics[width=0.9\textwidth]{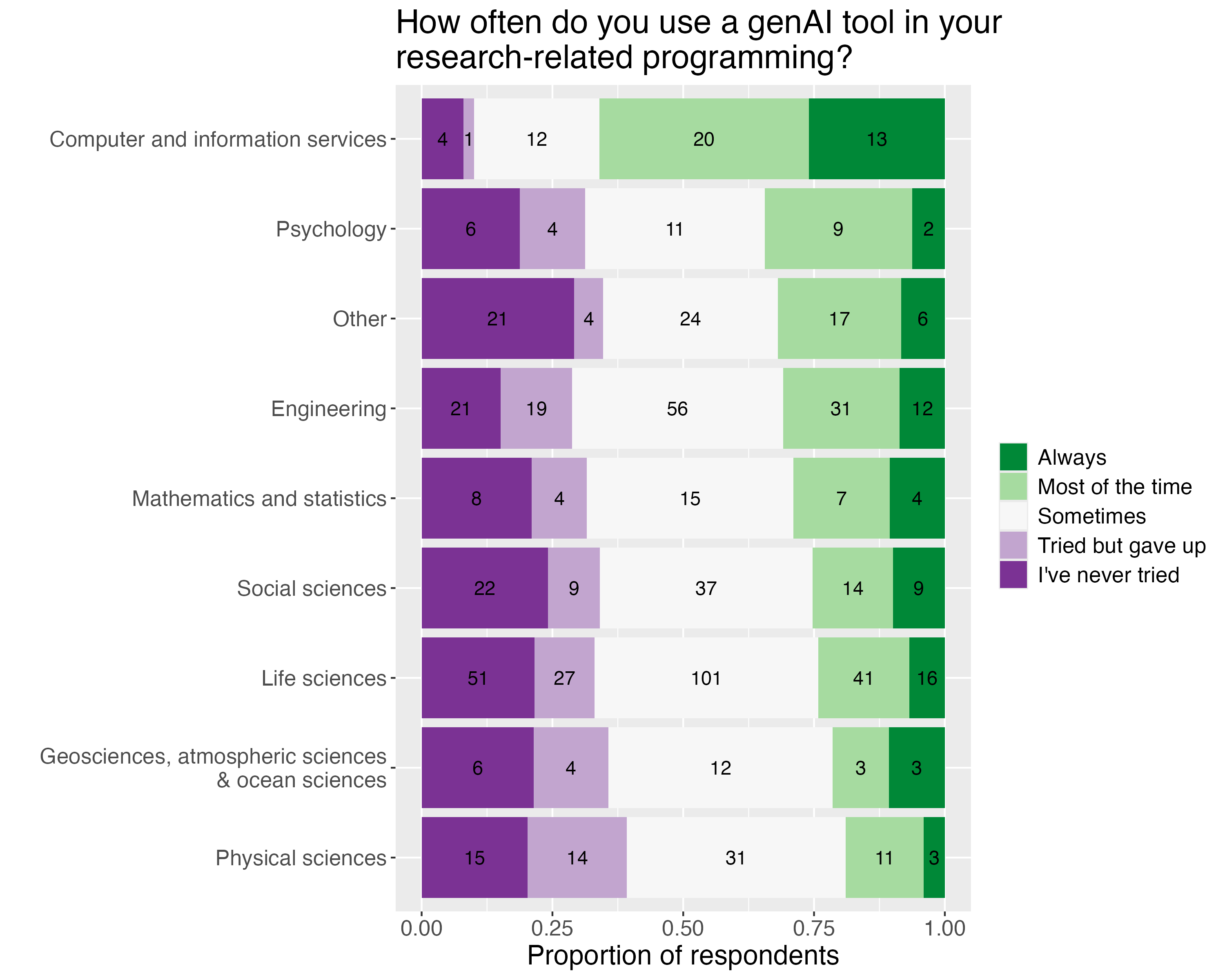}
\caption{Generative AI tool usage frequency for research-related programming by research area categories. Labels on bars indicate counts, and the $x$-axis indicates proportions. There were 760 responses to this question, and all respondents also provided a research area.}\label{fig:adoption_by_field}
\end{figure}

\paragraph{Difference by research area} 
There was an overall significant difference in GenAI usage frequency across disciplines ($F(8,751) = 4.22$, $p = 5.84e-05$). This appears driven mostly by Computer and Information Services, which showed a markedly higher proportion of frequent users than other disciplines (see Figure~\ref{fig:adoption_by_field}).   However, it is important to consider that the absolute number of respondents in some fields was much higher than in others (Life Sciences has 264 compared to Computer and Information Services' 57). As such, there are likely more users in absolute terms in fields with more moderate tool use. 

\paragraph{Difference by position} 

\begin{figure}[h]
\centering
\includegraphics[width=0.9\textwidth]{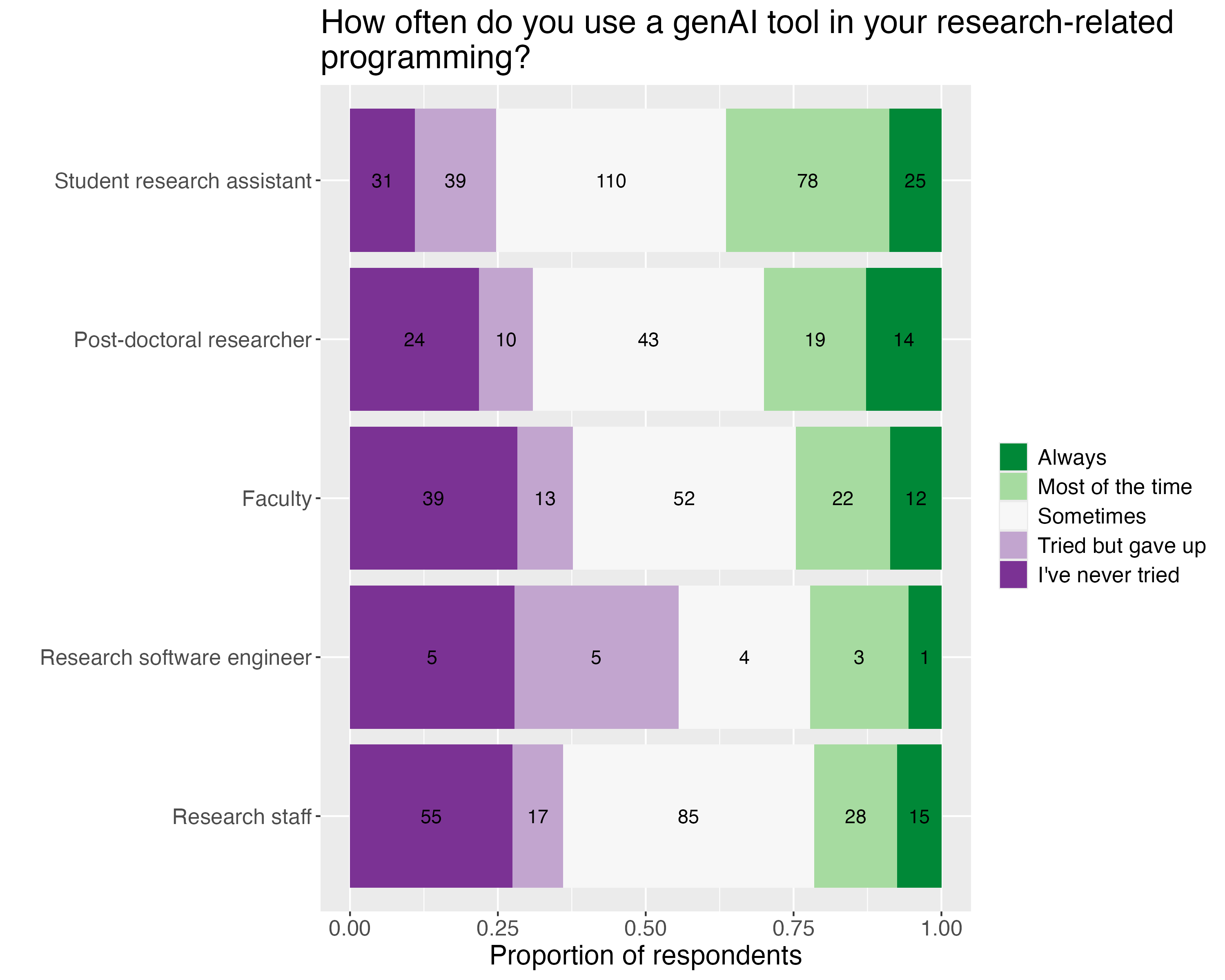}
\caption{Generative AI tool usage frequency for research-related programming by role. Labels on bars indicate counts.}
\label{fig:adoption_by_role}
\end{figure}

Usage frequency also varied significantly by respondents' position ($F(5,754) = 4.76$, $p = 0.0003$; Figure~\ref{fig:adoption_by_role}). Student research assistants showed the highest adoption rates. Research software engineers reported the lowest usage, though this difference should be interpreted cautiously because of the small sample size for this role (n=21). 

\paragraph{Years of programming experience}
Consistent with higher adoption among students, log-years of programming experience was inversely related to usage frequency (polyserial $r = -0.12$, $p = 0.001$; visualized in Figure~\ref{fig:adoption_by_experience}. Note that a polyserial correlation is used because usage frequency is assumed to be continuous, but discretized at the time of data collection by requiring participants to make a rating on a Likert-style scale \cite{Drasgow1986PolychoricCorrelations}). As such, less experienced programmers showed a modest but significant tendency to use genAI tools more frequently.

\paragraph{Gender}
Because it has been reported in at least one other study of scientific users of AI that women tend to adopt less frequently than men \cite{Chugunova2025WhoGermany}, we conducted an exploratory test of differences in adoption by gender. The omnibus effect of gender (modeling gender with the unordered categories "man", "woman", and "nonbinary") did not meet the threshold of significance ($F(2,741) = 2.68$, $p = 0.069$). However, there was a significant difference between women and men, with women tending to report less usage ($\beta = -0.20778$, SE = $0.09006$, $p = 0.0213$). This effect appears largely driven by an increased proportion of women who state they have tried programming with genAI tools (visualized in Figure~\ref{fig:adoption_by_gender}). 

\begin{table}
\caption{Primary tool (for adopters)}
\label{tab:primary_tool_choice}
\centering
\begin{tabular}[t]{lr}
\toprule
Tool & Count\\
\midrule
ChatGPT & 391\\
GitHub Copilot & 71\\
Google Gemini & 36\\
A custom tool provided by organization & 32\\
Claude & 26\\
Other (please write in) & 17\\
Microsoft Copilot & 11\\
Claude Code & 9\\
Cursor & 8\\
Perplexity & 8\\
\bottomrule
\end{tabular}
\end{table}

\subsection{What genAI tools are scientists who program using most?} 
Except for participants who indicated they had never programmed with genAI tools or had tried and given up, participants were asked what tools they had experimented with and what they considered their primary tool. Among 609 responses, ChatGPT was by far the most common (391/609, 64.2\%), followed distantly by GitHub Copilot (71/609, 11.7\%), Google Gemini (36/609, 5.9\%), and organization-provided custom tools (32/609, 5.3\%; Table~\ref{tab:primary_tool_choice} lists all responses). Note that our survey question did not distinguish between versions or "tiers" of each given tool (for example, distinguishing a professional ChatGPT subscription from free-tier), which would be necessary to fully document the capabilities of respondents' chosen tools.

General-purpose chat tools dominated over developer-specific tools: 472 respondents (77.5\%) selected ChatGPT, Google Gemini, Claude, Microsoft Copilot, or Perplexity, compared to just 88 (14.4\%) who chose dedicated programming tools like GitHub Copilot, Claude Code, or Cursor. This preference was reinforced by how users accessed their tools. When asked to indicate all access methods, 515 reported using a web browser, while only 100 indicated their tool was integrated into their programming environment as a plugin (some users of agentic developer tools like Cursor or Claude Code may have categorized these as desktop applications rather than IDE plugins, but this would not substantially change the overall pattern considering how few users these tools had).

Although we did not design our survey to explicitly address how genAI tools are chosen, we conducted an exploratory investigation into the factors that influence this decision. For this analysis, we excluded participants who indicated they used a custom tool provided by their organization, as we could not confidently classify these tools, and compare participants using a general-purpose tool (ChatGPT, Google Gemini, Claude, Microsoft Copilot, or Perplexity) versus a developer-specific tool (GitHub Copilot, Claude Code, or Cursor). Comparing these groups, users of developer tools tended to have more years of experience ($t(120.23) = 5.00$, $p = 1.93e-06$ for log$_{10}$-transformed years). The median user of a developer tool has 9 years of programming experience, compared to 6 for general-purpose tool users. There were also field-level differences: use of developer tooling was highest in Computer Science and Information Services, Mathematics and Statistics, and Physical Sciences, and lowest in Psychology, Social Sciences, and "Other" fields (Figure~\ref{fig:tool-choice-by-field}). This overall effect of the research area was significant, modeling the decision to use a developer tool instead of a general-purpose tool with a binomial generalized linear model ($\chi^2(8) = 25.68$, $p = 0.001$). Finally, there was also a marked gender difference: developer tool users were predominantly men (72.7\%), whereas general-purpose tool users were more evenly split by gender (50.9\% men, 47.0\% women, and 2.2\% non-binary or gender-diverse). The overall effect of gender on tool category was significant ($\chi^2(2) = 17.04$, $p = 0.0002$).

In summary, survey participants largely favored general-purpose tools like ChatGPT that are accessed via a browser. A minority of participants use developer tools like GitHub Copilot that are integrated into their development environment, and an even smaller number use largely agentic tools like Claude Code or Cursor. Users of targeted developer tools are more likely to be experienced developers, male, and working in Computer Science and Information Services. 

\subsection{What factors are associated with perceived productivity among scientists who program with genAI?}

We hypothesized that two factors in particular would be associated with greater perceived productivity with genAI tools: less programming experience and less use of software development practices (such as version control, code review, and code testing). 

To make this hypothesis testable, we need to operationalize measures of both perceived productivity and development practice adoption. To measure perceived productivity, we adapt the "SPACE" scale from a previous study of GitHub Copilot users \cite{Ziegler2022ProductivityCompletion}, which is designed to capture five theoretically-motivated \cite{Forsgren2021TheProductivity} aspects of developer productivity (\textbf{s}atisfaction and well-being, \textbf{p}erformance, \textbf{a}ctivity, \textbf{c}ommunication and collaboration, and \textbf{e}fficiency and flow). Full methodological details about this survey instrument are provided in Methods \ref{section:methods-space}. To measure adoption of software development practices, we adapted a survey instrument previously used to study adoption of industry-standard practices among scientific programmers \cite{Carver2022AStates}. 

We first report high-level patterns pertaining to the use of software development practices among survey respondents regardless of genAI tool use. We then report findings on perceived productivity and factors associated with it.

\subsubsection{Adoption of software development practices}

We assessed respondents' familiarity with and use of four practices common in professional software engineering: version control, code testing, code review, and continuous integration.  Participants first indicated whether they were familiar with each practice. Those reporting familiarity then rated their usage frequency on a 5-point scale. For code testing, we asked participants to rate their usage of three specific approaches—unit, regression, and system testing—as these were the most commonly adopted test types among scientific software developers in related work by Carver et al. \cite{Carver2022AStates}.

\begin{figure}[h]
\centering
\includegraphics[width=0.9\textwidth]{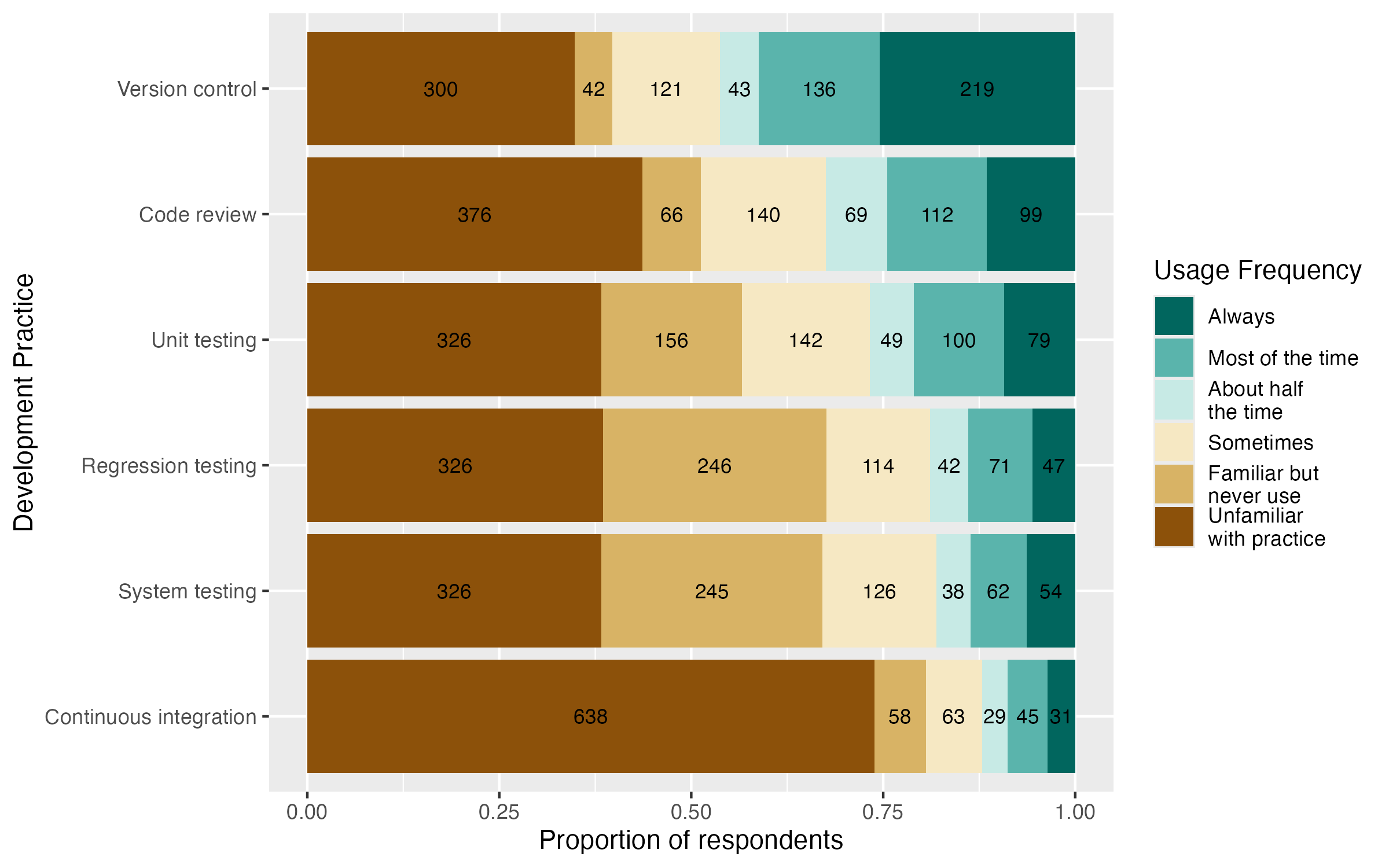}
\caption{Familiarity with and usage frequency of software development practices among survey respondents. Labels on bars indicate counts, and the $x$-axis indicates proportions. Note that respondents may skip survey items, so the number of total responses per practice can differ.}\label{fig:practice_adoption_counts}
\end{figure}

Overall, systematic adoption of development practices was modest (Figure~\ref{fig:practice_adoption_counts}). Version control was the most widely known and used: 567 respondents (65.4\%) reported familiarity, with 355 using it "most of the time" or "always." Code review was the next most common, followed by code testing. Testing practices showed particularly low adoption. Among the three test types we assessed, usage rates were consistently low: fewer than half of respondents familiar with each approach used it regularly. Continuous integration was the least familiar practice overall—only 228 respondents (26.3\%) had heard of it, and among those, just 76 used it more than half the time.

Most respondents reported familiarity with only a subset of these practices (median = 2), and nearly one-fifth (191) were unfamiliar with any. There were strong pairwise correlations between participants' ratings of their uses of each practice (polychoric $ r > 0.5$ for all pairs, see Figure~\ref{fig:dev_practices_correlation_matrix}; note that polychoric correlations are between two variables that are both discretized via Likert-style ratings \cite{Drasgow1986PolychoricCorrelations}). To summarize the overall adoption of development practices, we constructed a composite "development practice score" by averaging the ranked score associated with each practice. This composite variable was strongly correlated with each of the individual practice scores ($0.73 \leq  r \leq 0.89$, see Figure~\ref{fig:dev_practices_correlation_matrix}), suggesting it is a reasonable summary. 

Development practice scores were positively correlated with programming experience ($r = 0.29$, $p < 2.2e-16$), indicating that more experienced programmers were more likely to adopt systematic practices. Practice scores also varied significantly across disciplines ($F(8,857) = 6.53$, $p = 2.72e-08)$). Computer and Information Services researchers reported higher practice adoption than all other fields; this disciplinary difference persisted even after controlling for programming experience (model comparison is presented in Table~\ref{tab:dev_score_by_field}). 

\subsubsection{Code publishing and reuse}
Beyond formal development practices, we examined code sharing and reuse patterns, as these create opportunities for communal review that may surface problems. The distributions of responses to questions about code publishing, internal reuse, and external reuse are shown in Figure~\ref{fig:code_reuse}. Nearly 80\% (682/867 responses) of scientists published code associated with scientific work at least "sometimes", with about 25\% (211) doing so "most of the time" and 22.5\% (193) doing so "always". However, code publication differed from code reuse: when asked how often their code was reused within their own research group, the proportion of respondents answering "most of the time" or "always" dropped from 47.1\% to 35.6\% (205 "most of the time", 103 "always"). The proportion dropped even further for external reuse by scientists outside the respondents' research group, with just 9.5\% (82) responding "most of the time" (59) or "always" (23). The pattern of responses to these three questions indicates that while code publishing may be relatively common, it may not correspond closely to reuse.

\subsubsection{Factors related to perceived productivity}
We now turn to reporting the results of the perceived productivity survey items. Similar to Ziegler et al. \cite{Ziegler2022ProductivityCompletion}, we found overall high correlations between responses to the items on the SPACE perceived productivity questionnaire without structure that obviously supports a multifactorial model of perceived productivity (Figure~\ref{fig:space_correlations}). Therefore, we followed their analysis method and created a composite perceived productivity score by averaging responses to individual items (treating the ratings as numerical values on a 1 to 5 scale). This summary score was strongly correlated with each individual item ($0.68 \leq r \leq 0.80$). 

The perceived productivity score could range from 1 to 5, but tended toward the higher end with a mean of 3.9 ($\sigma = 0.70$) and median of 4. This trend is likely because the productivity questionnaire was only administered to respondents who indicated they program at least sometimes with a genAI tool, and we expect participants who did not perceive any productivity gains to be less likely to adopt a tool (and indeed, many non-adopters reported loss off productivity during trial periods, which is reported in the following section \ref{section:non-adoption}). Perceived productivity scores could be calculated for 623 respondents. 

\begin{figure}[h]
\centering
\includegraphics[width=0.9\textwidth]{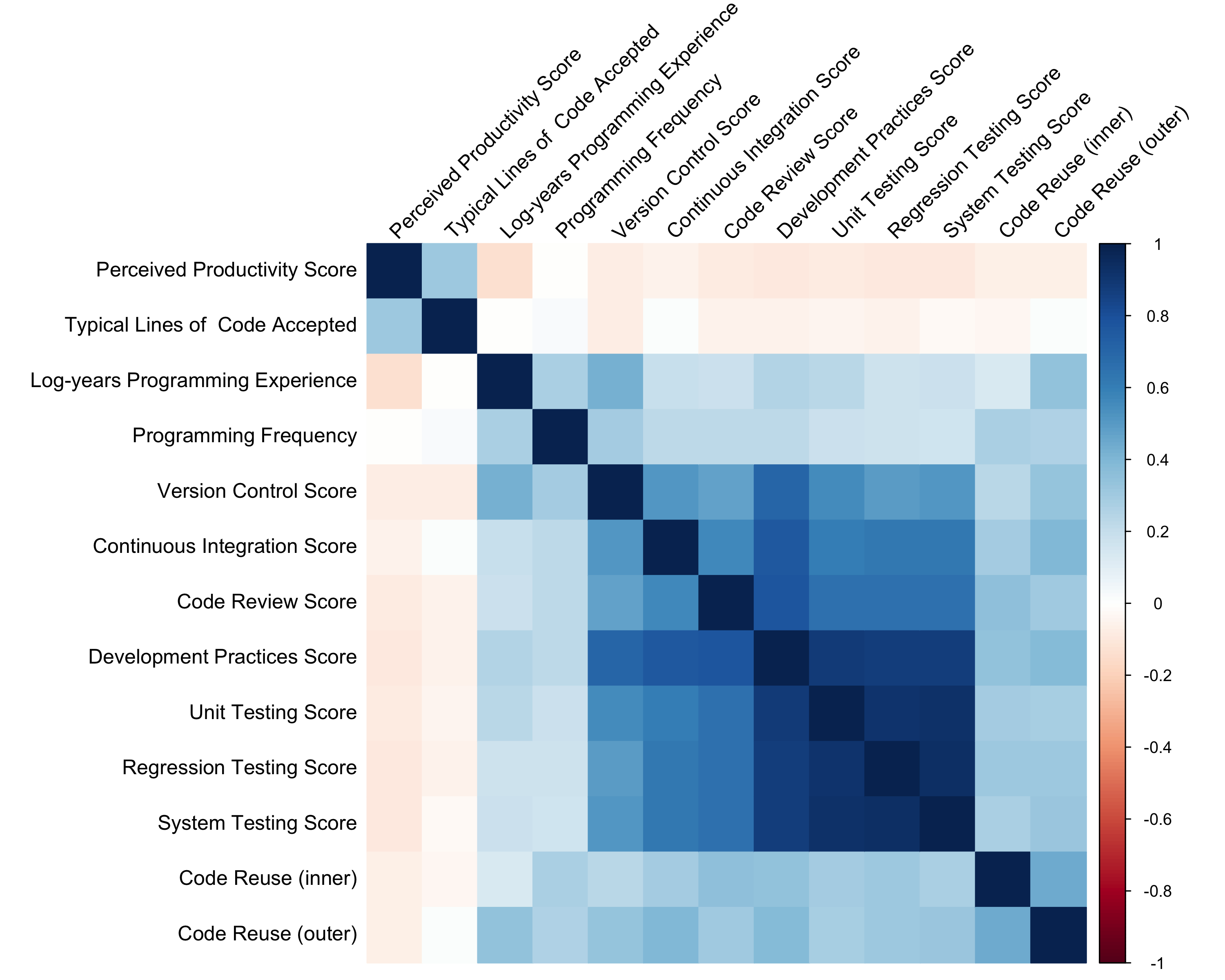}
\caption{Pairwise correlation matrix for perceived productivity score and variables related to development practices and programming experience. The matrix is ordered via hierarchical clustering.}\label{fig_productivity_corr}
\end{figure}

Pairwise correlations are presented for the perceived productivity score versus programming practice and development-related variables in \ref{fig_productivity_corr}, using polyserial and polychoric correlations as appropriate for correlations involving variables that are ordered factors (“Version Control Score”, “Code Review Score”, etc.). Perceived productivity was negatively associated with log-years of programming experience ($r = -0.14$, $p = 0.0009$) and with all the individual and composite development practice scores. We also checked whether perceived productivity differed by research area, but did not see evidence for an overall effect ($F(8,614) = 1.590$, $p = 0.125$). 

\begin{figure}
\centering
\includegraphics[width=0.9\textwidth]{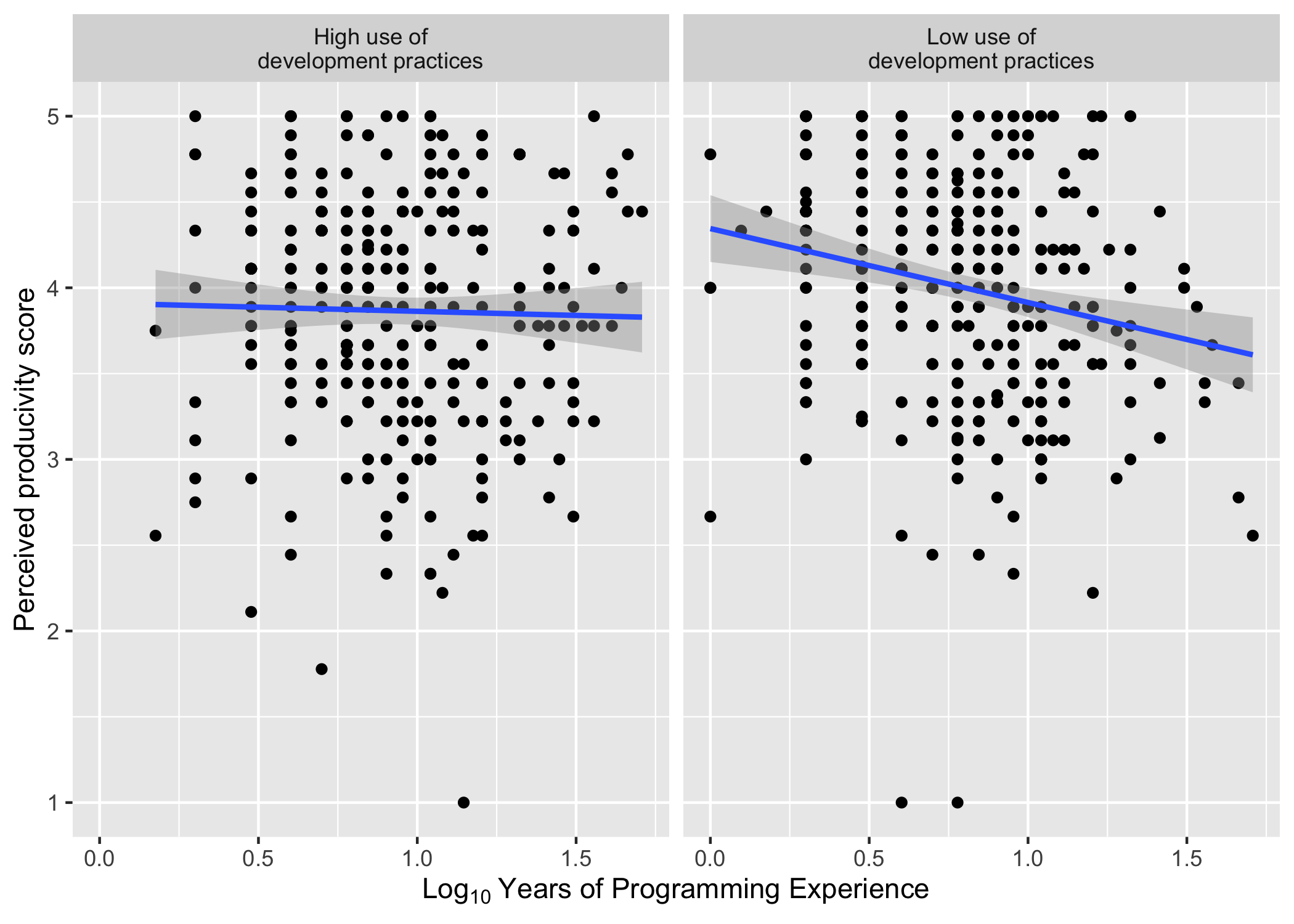}
\caption{Programming experience versus perceived productivity is shown faceted by development practice percentile. The sample is split on the median development practice score, and then a linear regression is fit for either split. Blue lines are lines of best fit for the data shown in the scatter plot of each pane with standard errors shaded in gray. Note that the median split is a simplification for visualization purposes only; in the main analysis via linear model, the development practice score is treated continuously. }\label{fig:space_dev_split}
\end{figure}

\begin{table}[!htbp] \centering 
  \caption{Model of perceived productivity} 
\label{tab:perceived-productivity-model} 
\begin{tabular}{@{\extracolsep{5pt}}lD{.}{.}{-3} } 
\\[-1.8ex]\hline 
\hline \\[-1.8ex] 
 & \multicolumn{1}{c}{\textit{Dependent variable:}} \\ 
\cline{2-2} 
\\[-1.8ex] & \multicolumn{1}{c}{Perceived productivity score} \\ 
\hline \\[-1.8ex] 
Development practice score & -0.217^{***} \\ 
  & (0.059) \\ 
  & \\ 
 $\log_{10}$-years programming experience & -0.554^{***} \\ 
  & (0.126) \\ 
  & \\ 
 Development practice score: $\log_{10}$-years programming experience & 0.207^{***} \\ 
  & (0.062) \\ 
  & \\ 
 Constant & 4.450^{***} \\ 
  & (0.108) \\ 
  & \\ 
\hline \\[-1.8ex] 
Observations & \multicolumn{1}{c}{602} \\ 
R$^{2}$ & \multicolumn{1}{c}{0.040} \\ 
Adjusted R$^{2}$ & \multicolumn{1}{c}{0.035} \\ 
Residual Std. Error & \multicolumn{1}{c}{0.659 (df = 598)} \\ 
F Statistic & \multicolumn{1}{c}{8.291$^{***}$ (df = 3; 598)} \\ 
\hline 
\hline \\[-1.8ex] 
\textit{Note:}  & \multicolumn{1}{r}{$^{*}$p$<$0.1; $^{**}$p$<$0.05; $^{***}$p$<$0.01} \\ 
\end{tabular} 
\end{table}

\begin{figure}[h]
\centering
\includegraphics[width=0.9\textwidth]{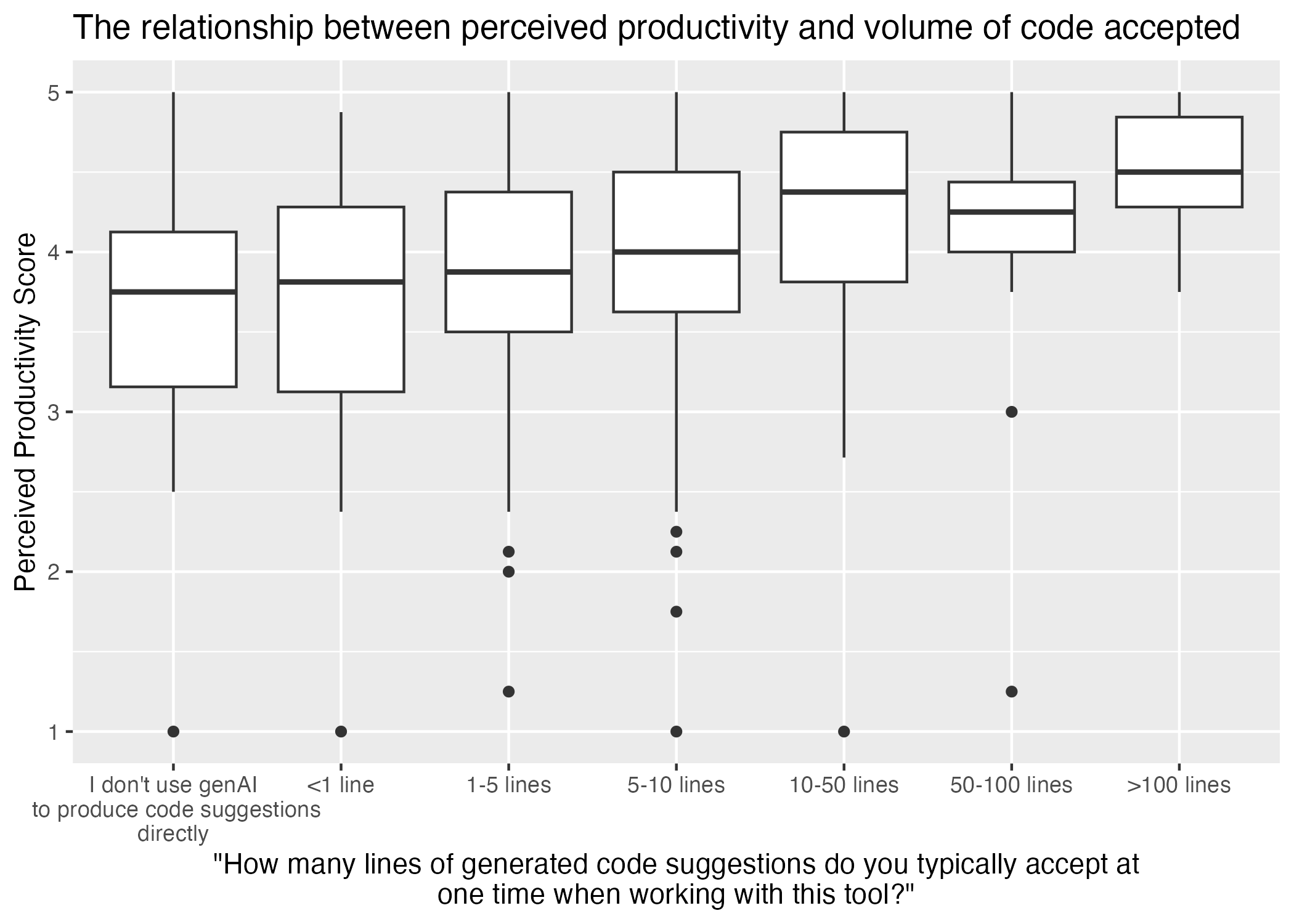}
\caption{Boxplots showing distribution of perceived productivity scores, split by response to the question on the $x$-axis label.}\label{fig_lines_accepted_productivity}
\end{figure}

One unexpected finding was the substantial relationship between perceived productivity score and answers to the question, “How many lines of generated code suggestions do you typically accept at one time when working with this tool?” (Figure~\ref{fig_lines_accepted_productivity}). In general, the more lines of code a respondent reported accepting at once, the more likely they were to report a high perceived productivity score. This relationship was significant (polyserial $r = 0.31$, $p=3.12e-17$) and explained about 9.6\% of variance in perceived productivity scores. Although small in absolute terms, this question explained more variance in perceived productivity than any other single factor studied. One possible explanation is that the SPACE questionnaire contains an item about producing more lines of code when using a genAI tool (see Table~\ref{tab:space_items}, "Activity"). As a robustness check, we recalculated the correlation with an average of responses to the SPACE scale excluding this item ($r = 0.28$, $p=1.01e-13$). Based on the persistent magnitude of the correlation, the relationship does not appear trivially explained by this item.

It is important to emphasize that scientists who report accepting a lot of generated code may still modify the code manually. In fact, a few open-ended responses to other questions describe generating large project skeletons with the intention of making substantial revisions. We do not know what percentage of accepted code corresponds to ultimately unmodified code.

Guided by our hypotheses about factors influencing perceived productivity, we modeled the perceived productivity score as a function of the composite development practice score and programming experience. We fit models with either additive or interacting predictors; an ANOVA comparison indicated better model fit with the interaction model ($F(1, 598) = 11.279$, $p=0.0008$). 

This fitted model is reported in Table~\ref{tab:perceived-productivity-model}.  We observe main effects of both development practice score and log-years of programming experience. The direction of these effects indicates that with increased adoption of both development practices and years of experience, there is an associated decrease in perceived productivity. There is also a significant interaction between the two factors, which is visualized in Figure~\ref{fig:space_dev_split}. For respondents who report relatively high use of development practices, there is a relatively flat relationship between programming experience and perceived productivity, whereas for respondents with lower usage of development practices, there is a more marked (inverse) relationship between years of experience and perceived productivity. One ad hoc explanation for this finding may be that in settings where scientists have little access to feedback about their code quality via development practices (like code review or testing), informal practices informed by years of experience guide scientists to exercise more critical reflection on generated code, which in turn moderates perceptions of productivity.  

We note also that the overall explanatory power of the model testing our hypothesis is low: $r^2=0.04$. It is surely possible to increase the explanatory power by adding more predictors---for example, adding a covariate for lines of code typically accepted (as a ranked factor) to the model in Table~\ref{tab:perceived-productivity-model} increases $r^2$ to 0.13 (adjusted $r^2 = 0.12$). However, an exhaustive search for the most predictive model is outside the goals of our study. 

To summarize, we see evidence that scientists with less programming experience and less adoption of a variety of development practices tend to rate themselves as more productive with genAI tools. The strongest single predictor of perceived productivity was reporting accepting longer blocks of code from genAI tools at once.

\subsection{What are reasons for non-adoption of genAI programming tools?}\label{section:non-adoption}
To address our final research question, respondents who indicated they had either not tried any tools or had tried but given up were directed to an open-ended response item asking what they attributed this decision to. We conducted a qualitative analysis of 210 responses to identify recurring themes (Methods~\ref{section:qualitative-methods}). The themes we identified are reported with examples and frequency counts in Table~\ref{tab:axial_codebook}. We briefly review the five most common themes, noting that responses could contain more than one.

\paragraph{Self-reliance} Many participants were concerned that using AI tools would prevent them from developing self-sufficiency as a programmer (45/210, or 21.4\% of responses). "I prefer to understand the code I am writing; I will not learn it well if the answer is generated for me." Some respondents acknowledged that this learning process was likely to involve a significant time investment, but they preferred to do so:  "I would rather spend time learning to code and write my own code rather than debugging AI-generated code."

\paragraph{Ethical concerns} An equally frequent theme (45/210, 21.4\%) was ethical concerns. Although ethical concerns are tallied as a group, there were several distinct subthemes, which are reported in Table~\ref{tab:ethics_subthemes}. First, ethical concerns frequently focused on the environmental impacts of training and using AI models, and often weighed potential individual benefits against the threat of broader harms: "The environmental and social impacts of the use of these tools isn't worth the small amount of time they would save me," wrote one respondent. 

Another prevalent ethical theme was distrust of AI as an industry, beyond the scope of considering a single product or model: "I believe that generative AI companies are engaged in highly immoral practices despite recognizing the negative impacts of their work," wrote one scientist. Others stated that "the training of AI is basically theft" or "regularly breaks copyright".

\paragraph{No demand} Some respondents reported they did not have enough programming work, or complex enough software, to warrant learning genAI tools (42/210, 20\%). For example, "I don't write enough for it to be worth my time." Reasons for not perceiving any demand for tools could include working on code that is very familiar, static ("I often am applying the same code to new datasets and don't need to make substantive changes"), or short ("I find that for small programs ($<100$ lines), it is easier to just code it myself than bother with AI."). 

\paragraph{Inefficiency} Some respondents did not perceive time savings with the tools they had tried (41/210, 19.5\%). "It takes more time to figure out what is wrong when I don't write the code," wrote one respondent. Another respondent described a time-consuming experience trying to pinpoint the origin of problems in generated code: "It's rare to get a piece of code out of these systems that is usable. Even small code like, say, generating a table for a latex doc inevitably has bugs. Once ChatGPT create an API call that simply did not exist.  I wasted hours figuring out what the heck was going on."

\paragraph{Accuracy} Another common theme was encountering issues with the reliability of tools to produce accurate code generations or answers to coding-related questions (40/210; 19\%). One scientist who had tried, but given up working with ChatGPT wrote, "I found it unable to generate testable, correct code for my purposes". Another compared reliability negatively to a search engine: "The hallucinations in the tools were too annoying. At least with google I get accurate information."

\setlength{\tabcolsep}{4pt}
\renewcommand{\arraystretch}{1.5} 

\begin{longtable}{p{0.18\textwidth} p{0.30\textwidth} p{0.42\textwidth} r}
\caption{Codebook for reasons for non-adoption of genAI tools}\label{tab:axial_codebook}\\
\toprule
Label & Description & Example & Count\\
\midrule
\endfirsthead

\multicolumn{4}{l}{\small\textit{(continued from previous page)}}\\
\toprule
Label & Description & Example & Count\\
\midrule
\endhead

\midrule
\multicolumn{4}{r}{\small\textit{(continued on next page)}}\\
\bottomrule
\endfoot

\bottomrule
\endlastfoot
Self-Reliance & Concern that the usage of AI prevents the development of programming or research skills 
& ``I’ve occasionally used them when I’m stuck, but I really want to improve my coding skills and I think the best way to do so is to write the code myself and work through bugs.'' 
& 45\\

Ethics & Ethical concerns about AI technologies and as an industry, including environmental impacts, company practices, and broader societal risks 
& 
``I see serious ethical concerns in the training of these models and believe that generative AI companies are engaged in highly immoral practices.'' 
& 45\\

No Demand & Expresses that tools would be superfluous for their use case 
& ``Processes are basic and familiar at this point so have yet to feel the need.'' 
& 42\\

Inefficiency & Tools decrease productivity or fail to save time 
& ``While genAI can start a code from zero, it still needs a lot of rewriting to fit my purpose. I found that tuning the genAI codes is not actually saving much time for me.'' 
& 41\\

Accuracy & Concern about the accuracy or reliability of generated code and information
&  ``The code that AI tools produce is horrible – full of bugs and logical mistakes.'' 
& 40\\

Existing Tool & Support needs are met by an existing workflow or set of tools
& ``I have mostly used code for pretty straightforward analysis and I use in-program help commands and published resources for help.'' 
& 21\\

Unfamiliar & Low awareness of tools available or how to use them 
& ``Not familiar/comfortable enough with them.'' 
& 19\\

Privacy & Concerned about compromising sensitive information 
& ``Not familiar with the tools and have concerns about data privacy.'' 
& 10\\

Enjoyment & Avoids tools because they find personal satisfaction and fulfillment in their current programming practice
& ``I enjoy the process of writing code and I'm not eager to offload it to genAI.'' 
& 9\\

Time Investment & Learning a new tool requires too much time 
& ``I see potential but do not feel I have the time to learn the new tool.'' 
& 9\\

Disinterest & Expresses broad disinterest
& ``I'm not particularly interested in genAI overall.'' 
& 7\\

Selective Use & Uses tools only in narrow cases 
& ``I only use it if I want to Google something specific.'' 
& 7\\

Niche & Tool performance on specific research computing needs is poor
& ``They're not sufficiently trained. The packages I use are too niche.'' 
& 4\\

Tool Integration & Could not integrate AI tool into existing workflow 
& ``I couldn't get Copilot to work in VS Code in vim mode.'' 
& 2\\

Access & Cost or access barriers 
& ``Sometimes it is difficult or the resources are not free to use.'' 
& 1\\

\end{longtable}

\section{Discussion}\label{discussion}
Within our non-random survey sample of scientific programmers, GenAI tool adoption
for research-related programming is common but heterogeneous. Adoption is highest
among junior scientists and less experienced programmers, with markedly higher adoption in Computer \& Information Services compared to other fields. As of the summer 2025 data collection period, respondents overwhelmingly prefer general-purpose, proprietary conversational tools like ChatGPT accessed through web browsers over developer-specific tools integrated into programming environments. About a quarter of respondents did not regularly use genAI coding tools, and common reasons included concerns about hindering skill development, ethical issues (particularly environmental impacts), lack of perceived need, lack of productivity gains during trials, and accuracy problems.

Perceived productivity is associated with programming inexperience and limited
adoption of development practices like code testing or review. Our work replicates and extends the findings of Ziegler et al. \cite{Ziegler2022ProductivityCompletion}, who showed that perceived productivity with GitHub Copilot was higher among more junior programmers, to a new sample with more varied tool choices. Our findings also suggests that use of development practices could moderate perceptions of productivity: there is a stronger relationship between programming experience and perceived productivity when the use of development practices is relatively low.  The strongest single predictor of perceived productivity among our sample is the
volume of code accepted at once, with users accepting $>$100 lines reporting the
highest productivity. Together, we interpret these findings to suggest that scientific programmers may gauge productivity primarily by code generation rather than validation.

We emphasize that findings should be interpreted in context of our study design: we used a non-random, mostly academic, and heavily U.S.-centric sample, limiting generalizability to non-academic researchers and scientists in the global south. Recruitment through scientific software communities likely under-represents more casual programmers. Because our survey lacks direct measures of code quality or developer activity, we cannot definitively determine how perceived productivity relates to the quality of software that is actually being produced by respondents. The SPACE scale has theoretical and experimental precedents~\cite{Forsgren2021TheProductivity, Ziegler2022ProductivityCompletion}, but its reliability and construct validity have not been rigorously evaluated. The correlation structure we observed between survey items suggests the five theoretical dimensions may not correspond to five distinct factors---as in Ziegler et al.~\citep{Ziegler2022ProductivityCompletion}, we observed strong cross-dimension correlations and therefore use a composite score during modeling. In practical terms, this means that a considerably abbreviated scale may capture much of the same variance as the full scale in follow up studies. With regard to the theory developer productivity, though, the five theorized dimensions may warrant re-inspection based on a factor analysis. Lastly, self-reported usage of development practices is a coarse description: a scientist might report that they "use" version control sometimes, but not particularly adhere to practices beyond uploading files to GitHub. We lack any previous measures that would contextualize whether respondents are prone to over- or under-estimating their use of such practices.

Our finding that greater perceived productivity is associated with
inexperience and low use of development practices could have several explanations. One
is automation bias, in which code is accepted uncritically due to a lack of scaffolding for validation---a pattern documented broadly in human-computer
interaction~\cite{Goddard2012AutomationMitigators} and specifically for genAI coding
tools~\cite{Prather2024TheProgrammers, OBrien2025HowProgram,
MoradiDakhel2023GitHubLiability}. The association between high perceived
productivity and accepting the most generated code could be consistent with this
explanation. If automation bias is widely occurring, then we expect that the quality of scientific software may be broadly at risk unless there is a rise in compensatory validation work. Alternatively, beginner programmers and those who do not follow standard development practices may indeed be well-calibrated to how genAI is improving their scientific programming work. These scientists may differ in important ways from scientists who adhere to any development practices, which may be insufficiently distinguished by our survey. We also cannot rule out that very long code generations could be frequently used largely as scaffolding, with the intention to make substantial modifications later (in fact, we are sure this occurs in at least some cases, based on open-ended comments left by a few respondents). 

Our investigation assumes that development practices like testing, code review, and
version control support scientific code validity~\cite{Wilson2017GoodComputing,
Carver2022AStates, Eisty2025TestingTools, Vogel2019ChallengesScience,
ArifulIslamMalik2025PeerPerspective}. However, exploratory programming with an
emphasis on flexibility~\cite{BethKery2017ExploringProgramming} is an essential part
of some scientific work~\cite{Sutherland-Keller2025ResearchCosmology}, and
scientists have been observed to engage in testing behaviors without following industry norms~\cite{Kelly2011ScientificDimensions, Kelly2015ScientificSoftware,
Paine2017WhoVisualizations}. Capturing these non-formalized validation activities in
survey and behavioral measures will be important, and the absence of measures targeting these activities may be one reason that the majority of variance in perceived productivity is unexplained by the factors measured in our current study---a pattern also seen in Ziegler et al.~\cite{Ziegler2022ProductivityCompletion} (though this may alternatively reflect the scale's imperfect reliability).

No single development practice was an especially strong predictor of perceived productivity, suggesting a non-specific effect. We also observed similar correlations between perceived productivity and both internal and external code reuse
(Figure~\ref{fig_productivity_corr}), which suggests that if development practices
causally moderate productivity perceptions, the mechanism may be that increased
scrutiny---whether through testing, external reuse, or code review---provides feedback about code that would otherwise be absent. Testing this causal account could be accomplished via intervention studies or mediation modeling. 

Another potentially important direction for future research and practice pertains to tool choice. The majority of respondents use general-purpose conversational tools not specialized for software development, replicating a finding from a 2024 single-institution study by one of the investigators \cite{OBrien2025HowProgram}; notably, the proportion using these tools appears stable across the two surveys despite the year separating them. Many scientists are therefore using tools that \textit{happen} to support programming rather than tools designed for it. This may not be problematic---conversational interfaces may simply be more compatible with many scientists' existing workflows---but it raises questions about what capabilities go unused. This matter may be especially pressing given recent advances in the capabilities of agentic AI tools for programming (such as Claude Code or Cursor), which had low adoption in our sample. The predominance of proprietary tools also creates institutional risk: universities have little leverage over these external dependencies, and the closed nature of these systems limits how we can study their usage at scale. Understanding what drives scientific programmers toward proprietary tools---and what would need to change for open-source alternatives to compete---may be a consequential matter for the long-term sustainability of scientific software.

\backmatter

\bmhead{Supplementary information}
The survey questionnaire is provided as supplementary information. 

\bmhead{Acknowledgments}
This work is supported by a grant from the Alfred P. Sloan foundation to G.O.

\section*{Declarations}
\subsection*{Code Availability}
The code used to conduct all quantitative analyses presented here is available at \url{https://github.com/elleobrien/survey_of_scientific_programmers_genai}.

\subsection{Author contributions}
G.O. contributed to survey design, data collection, all data analysis, and writing. A.P. contributed to survey testing, qualitative analysis and writing. N.E. and J.C. contributed to survey design and writing. 

\subsection{Funding}
G.O. was funded by a grant from the Alfred P. Sloan Foundation.

\subsection*{Ethics declarations} 
\subsubsection*{Competing interests}
The authors declare no competing interests.

\begin{appendices}

\setcounter{table}{0}
\renewcommand\theHtable{Appendix.\thetable}

\renewcommand{\thefigure}{A\arabic{figure}}
\renewcommand{\theHfigure}{A\arabic{figure}}
\setcounter{figure}{0}

\section{Demographics}\label{sec:demographics}
\begin{figure}[hbt!]
\centering
\includegraphics[width=0.9\textwidth]{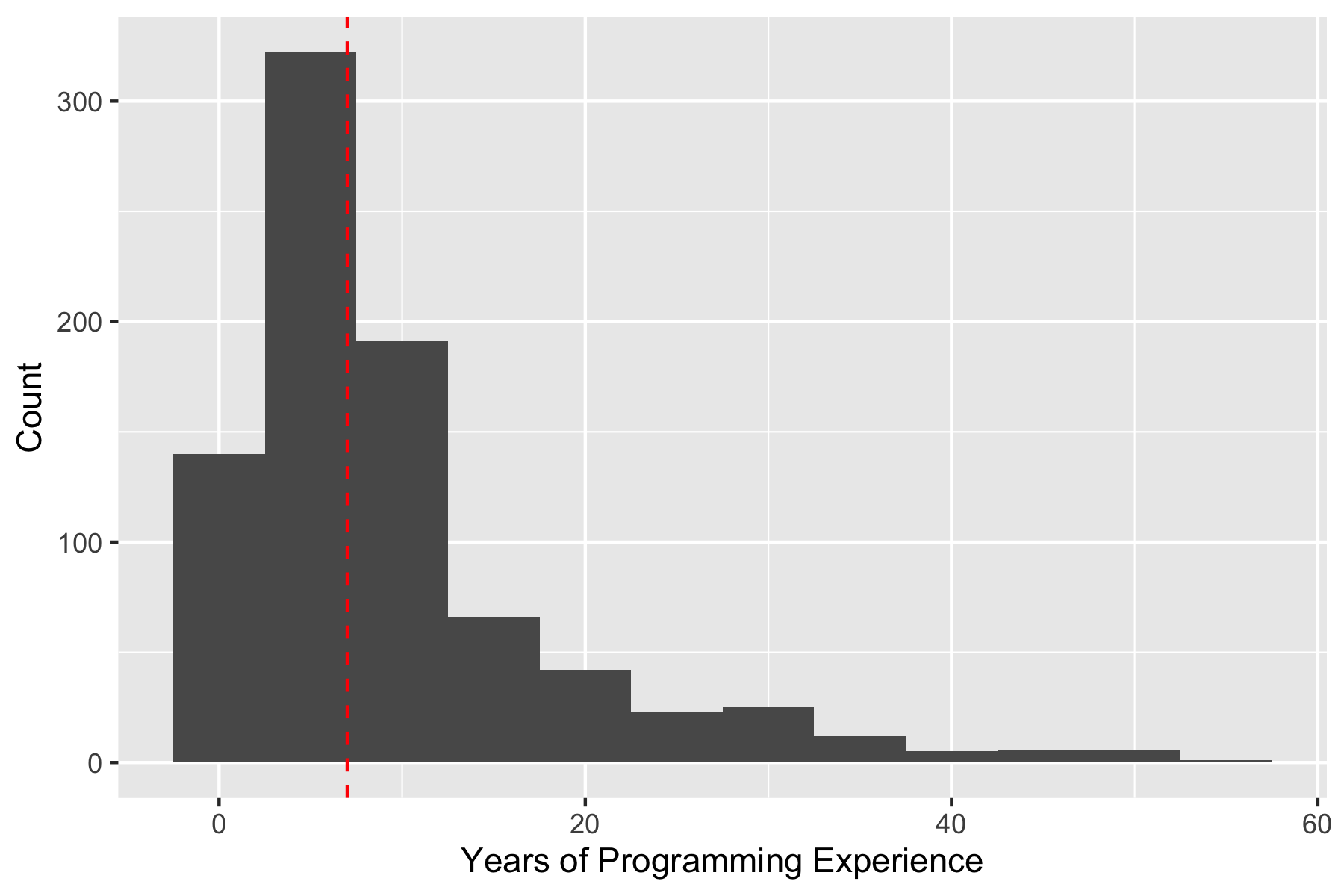}
\caption{Survey respondents' self-reported years of programming experience. Median is represented with dashed vertical line.}\label{fig:years_program_exp}
\end{figure}


\begin{table}[hbt!]
\caption{Programming languages reported by survey respondents}
\label{tab:programming_languages}
\begin{tabular}{lr}
\toprule
Language & Count\\
\midrule
Python & 549\\
R & 474\\
MATLAB & 257\\
Bash & 165\\
Other & 135\\
C++ & 120\\
Stata & 75\\
JavaScript & 55\\
C & 50\\
FORTRAN & 49\\
Java & 28\\
Julia & 19\\
Rust & 6\\
\bottomrule
\end{tabular}

\end{table}

\clearpage

\section{Adoption}\label{sec:gender_differences}

\begin{figure}[hbt!]
\centering
\caption{Log-years of programming experience, split by genAI usage.}\label{fig:code_reuse}
\label{fig:adoption_by_experience}
\includegraphics[width=0.9\textwidth]{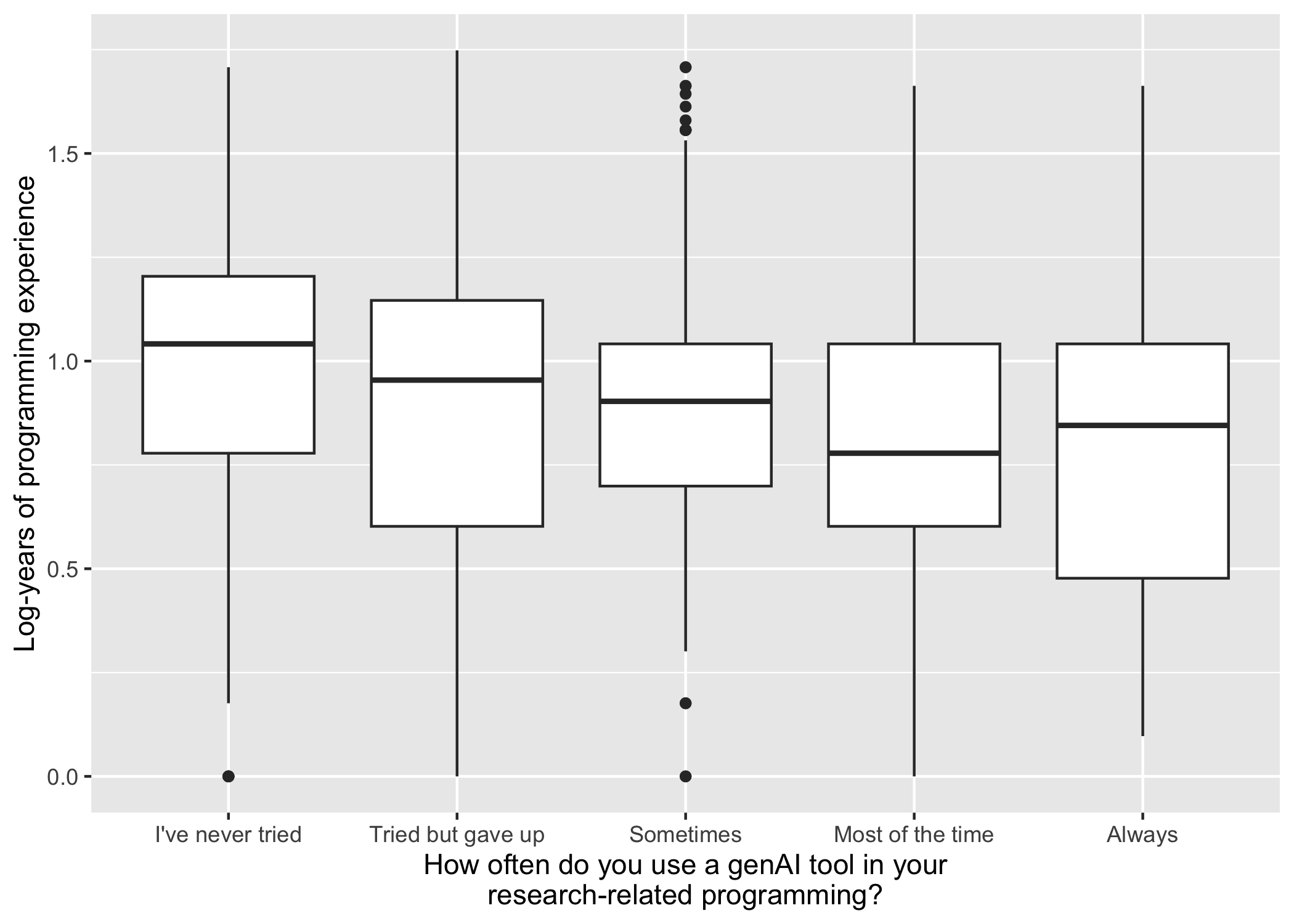}

\end{figure}

\begin{figure}[hbt!]
\centering
\includegraphics[width=0.9\textwidth]{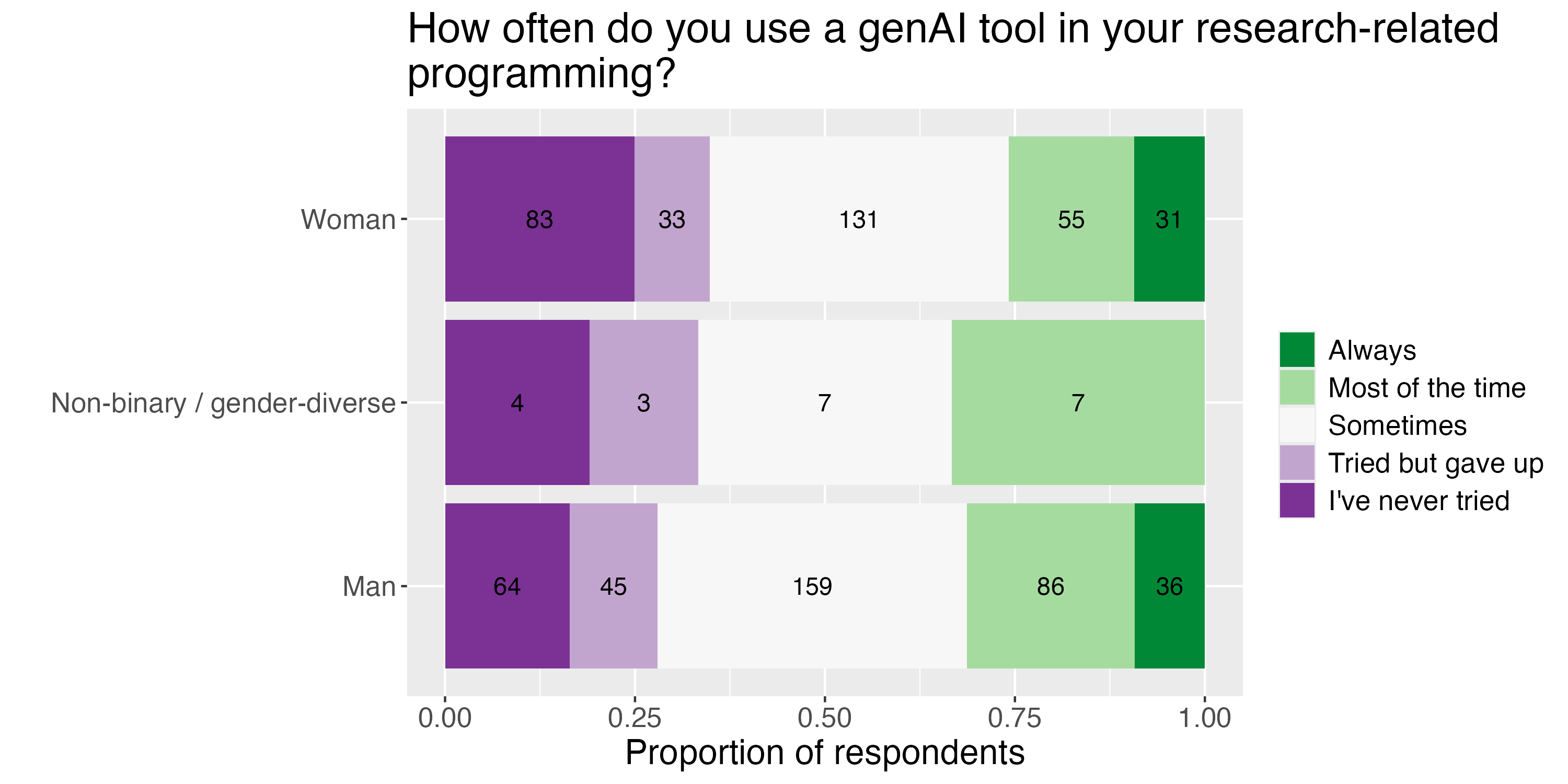}
\caption{Differences in adoption of genAI tools for programming by gender. Data is excluded for 15 participants who declined to provide their gender. Numbers on bars indicate absolute counts.}\label{fig:adoption_by_gender}
\end{figure}

\begin{table}[!htbp] \centering 
  \caption{Adoption and gender} 
  \label{tab:adoption_by_gender} 
\begin{tabular}{@{\extracolsep{5pt}}lD{.}{.}{-3} } 
\\[-1.8ex]\hline 
\hline \\[-1.8ex] 
 & \multicolumn{1}{c}{\textit{Dependent variable:}} \\ 
\cline{2-2} 
\\[-1.8ex] & \multicolumn{1}{c}{as.numeric(genai\_tool\_freq)} \\ 
\hline \\[-1.8ex] 
 genderNon-binary / gender-diverse & -0.152 \\ 
  & (0.270) \\ 
  & \\ 
 genderWoman & -0.208^{**} \\ 
  & (0.090) \\ 
  & \\ 
 Constant & 2.962^{***} \\ 
  & (0.061) \\ 
  & \\ 
\hline \\[-1.8ex] 
Observations & \multicolumn{1}{c}{744} \\ 
R$^{2}$ & \multicolumn{1}{c}{0.007} \\ 
Adjusted R$^{2}$ & \multicolumn{1}{c}{0.005} \\ 
Residual Std. Error & \multicolumn{1}{c}{1.207 (df = 741)} \\ 
F Statistic & \multicolumn{1}{c}{2.684$^{*}$ (df = 2; 741)} \\ 
\hline 
\hline \\[-1.8ex] 
\textit{Note:}  & \multicolumn{1}{r}{$^{*}$p$<$0.1; $^{**}$p$<$0.05; $^{***}$p$<$0.01} \\ 
\end{tabular} 
\end{table} 

\clearpage
\section{Programming practices}

\begin{figure}[hbt!]
\centering
\includegraphics[width=0.9\textwidth]{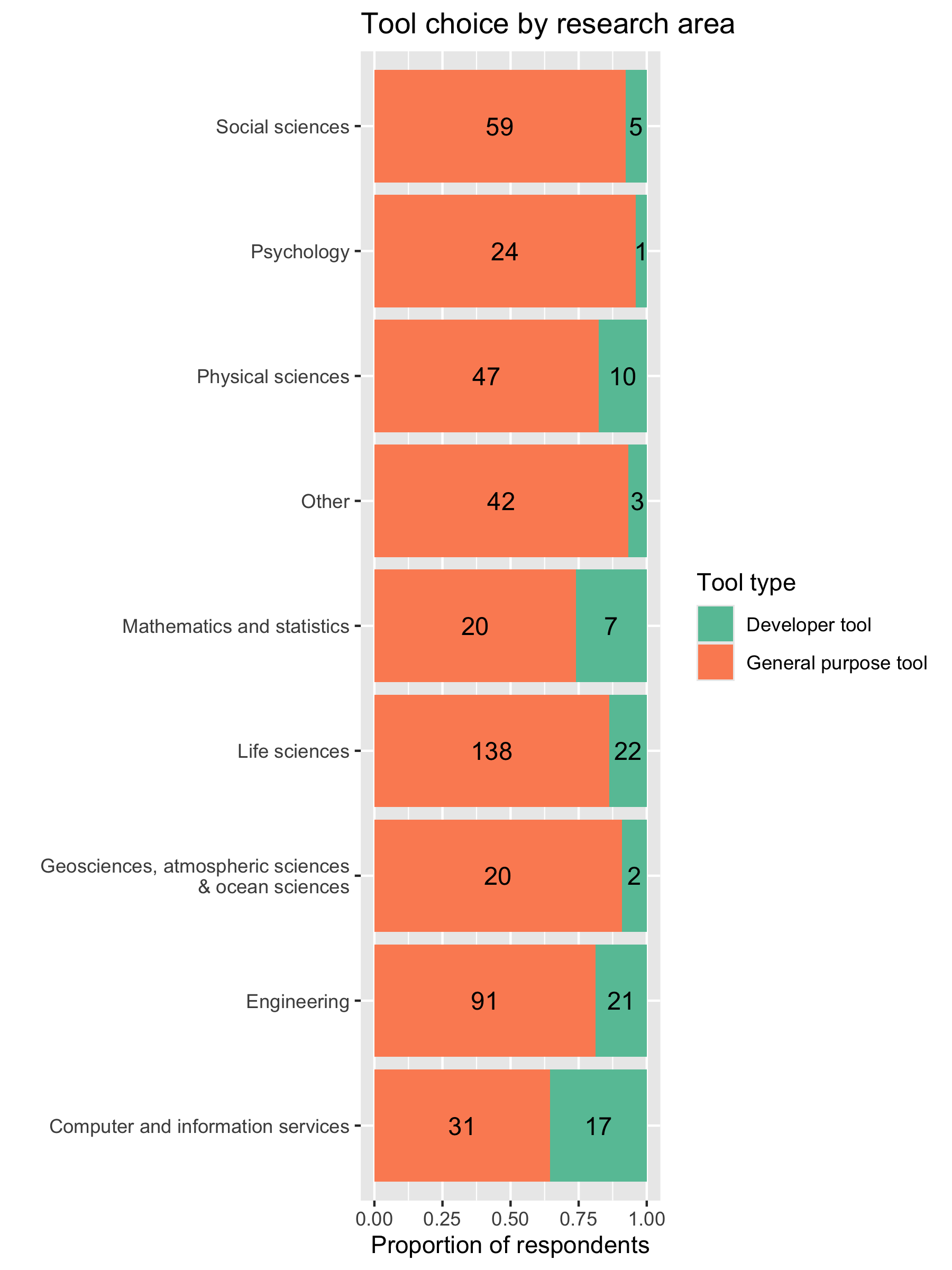}
\caption{Field-level differences in adoption of genAI tools in two categories, ``general purpose'' or ``developer specific''. General purpose tools are defined as ChatGPT, Claude, Microsoft Copilot, Google Gemini, and Perplexity. Developer specific tools are defined as GitHub Copilot, Claude code, and Cursor. For this exploratory analysis, participants who could not be clearly assigned to either category based on their primary tool choice (i.e., they selected ``a custom tool provided by organization'' or ``other'') are not included.}\label{fig:tool-choice-by-field}
\end{figure}

\begin{figure}[hbt!]
\centering
\includegraphics[width=0.9\textwidth]{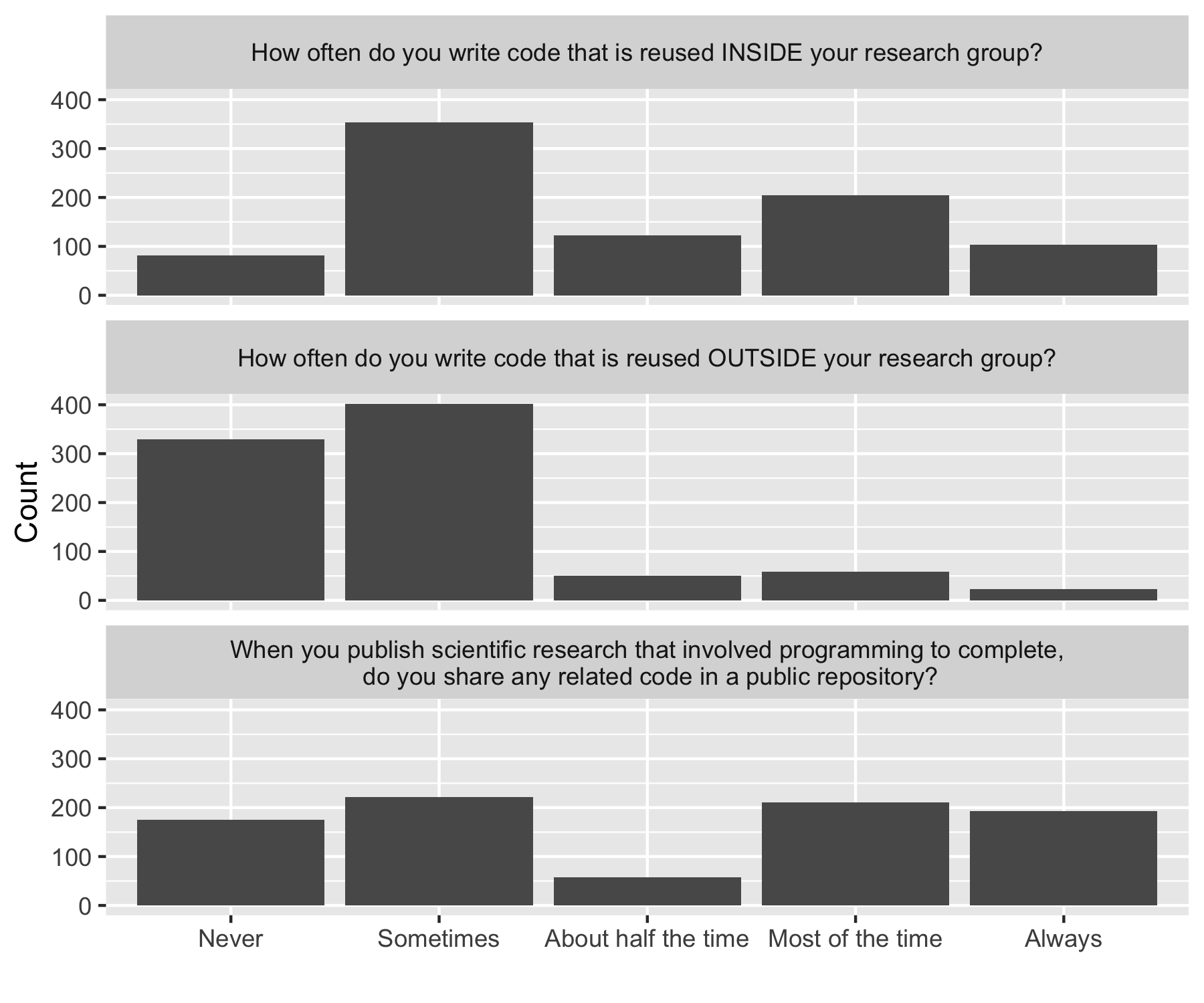}
\caption{Distribution of responses to questions about code publishing and reuse.}\label{fig:code_reuse}
\end{figure}

\begin{figure}[h]
\centering
\includegraphics[width=0.9\textwidth]{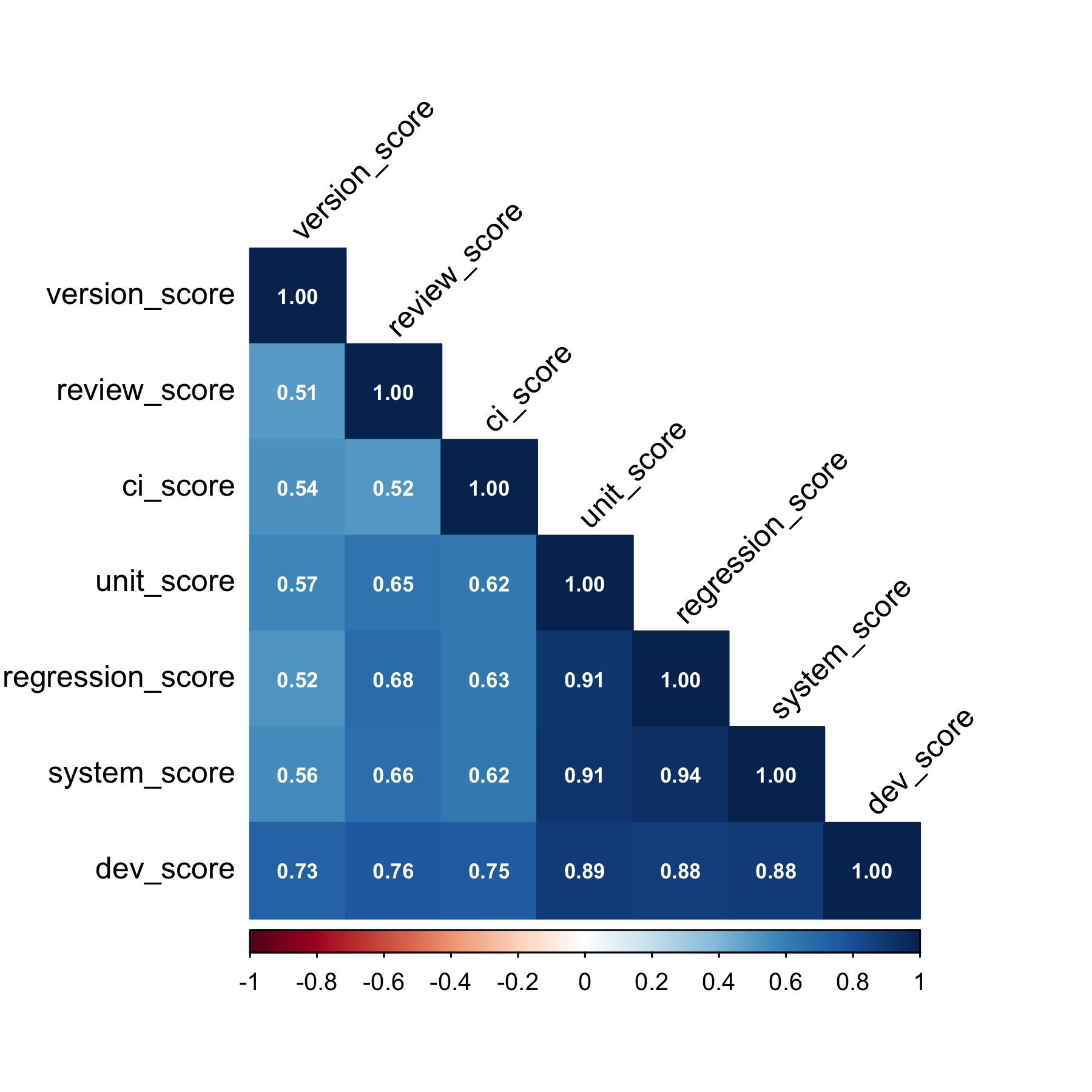}
\caption{Pairwise polychoric correlations between scores for 6 development practices and the composite development score \texttt{dev\_score}. }\label{fig:dev_practices_correlation_matrix}
\end{figure}

\begin{table}[!htbp] \centering 
  \caption{Development score differences by field. Note that Computer and information services category is the baseline category against which other fields are compared.} 
  \label{tab:dev_score_by_field} 
\begin{tabular}{@{\extracolsep{5pt}}lD{.}{.}{-3} D{.}{.}{-3} } 
\\[-1.8ex]\hline 
\hline \\[-1.8ex] 
 & \multicolumn{2}{c}{\textit{Dependent variable:}} \\ 
\cline{2-3} 
\\[-1.8ex] & \multicolumn{2}{c}{dev\_score} \\ 
\\[-1.8ex] & \multicolumn{1}{c}{(1)} & \multicolumn{1}{c}{(2)}\\ 
\hline \\[-1.8ex] 
 Field: Engineering & -0.902^{***} & -0.907^{***} \\ 
  & (0.193) & (0.188) \\ 
  & & \\ 
 Field: Geosciences, atmospheric sciences\& ocean sciences & -0.629^{**} & -0.681^{**} \\ 
  & (0.274) & (0.266) \\ 
  & & \\ 
 Field: Life sciences & -1.259^{***} & -1.185^{***} \\ 
  & (0.183) & (0.179) \\ 
  & & \\ 
 Field: Mathematics and statistics & -0.967^{***} & -1.169^{***} \\ 
  & (0.258) & (0.249) \\ 
  & & \\ 
 Field: Other & -0.975^{***} & -0.867^{***} \\ 
  & (0.219) & (0.213) \\ 
  & & \\ 
 Field: Physical sciences & -1.004^{***} & -1.000^{***} \\ 
  & (0.213) & (0.208) \\ 
  & & \\ 
 Field: Psychology & -1.165^{***} & -1.028^{***} \\ 
  & (0.258) & (0.250) \\ 
  & & \\ 
 Field: Social sciences & -0.917^{***} & -1.004^{***} \\ 
  & (0.206) & (0.201) \\ 
  & & \\ 
 Log-years programming experience&  & 1.055^{***} \\ 
  &  & (0.122) \\ 
  & & \\ 
 Constant & 2.492^{***} & 1.544^{***} \\ 
  & (0.166) & (0.199) \\ 
  & & \\ 
\hline \\[-1.8ex] 
Observations & \multicolumn{1}{c}{866} & \multicolumn{1}{c}{837} \\ 
R$^{2}$ & \multicolumn{1}{c}{0.057} & \multicolumn{1}{c}{0.136} \\ 
Adjusted R$^{2}$ & \multicolumn{1}{c}{0.049} & \multicolumn{1}{c}{0.127} \\ 
Residual Std. Error & \multicolumn{1}{c}{1.253 (df = 857)} & \multicolumn{1}{c}{1.192 (df = 827)} \\ 
F Statistic & \multicolumn{1}{c}{6.534$^{***}$ (df = 8; 857)} & \multicolumn{1}{c}{14.489$^{***}$ (df = 9; 827)} \\ 
\hline 
\hline \\[-1.8ex] 
\textit{Note:}  & \multicolumn{2}{r}{$^{*}$p$<$0.1; $^{**}$p$<$0.05; $^{***}$p$<$0.01} \\ 
\end{tabular} 
\end{table} 

\clearpage

\section{SPACE questionnaire}

\begin{table}[htbp]
\centering 
\caption{The SPACE questionnaire and aspects of productivity measured.}
\label{tab:space_items}
\begin{tabular}{p{7cm}p{4cm}c}
\toprule
\textbf{Survey statements} & \textbf{Productivity aspect} & \textbf{Code} \\
\midrule
``I am more productive when using this tool'' & Global productivity & \verb|space_global| \\
\midrule
``I find myself less frustrated during coding sessions when using this tool.'' & \multirow{2}{4cm}{Satisfaction and well-being} & \verb|space_s1| \\
\midrule

``While working with an unfamiliar language, I make progress faster when using this tool.'' & & \verb|space_p1| \\
\cmidrule{1-1}
``The code I write using this tool is better than the code I would have written without it.'' & \multirow{-2}{4cm}{Performance} & \verb|space_p2| \\
\midrule
``I produce more lines of code with this tool than I would have without it.`` & Activity & \verb|space_a1|\\
\midrule

``I learn from the responses this tool gives me.'' & & \verb|space_c1|\\
\cmidrule{1-1}
``I am better able to understand other people's code with this tool.'' & \multirow{-2}{4cm}{Communication and collaboration} & \verb|space_c2|\\
\midrule

``I spend less time searching for information or examples when using GitHub Copilot.'' & & \verb|space_e1|\\
\cmidrule{1-1}
``I complete programming tasks faster when using this tool.''' & \multirow{-2}{4cm}{Efficiency and flow} & \verb|space_e2| \\

\bottomrule
\end{tabular}
\end{table}

\begin{figure}[h]
\centering
\includegraphics[width=0.9\textwidth]{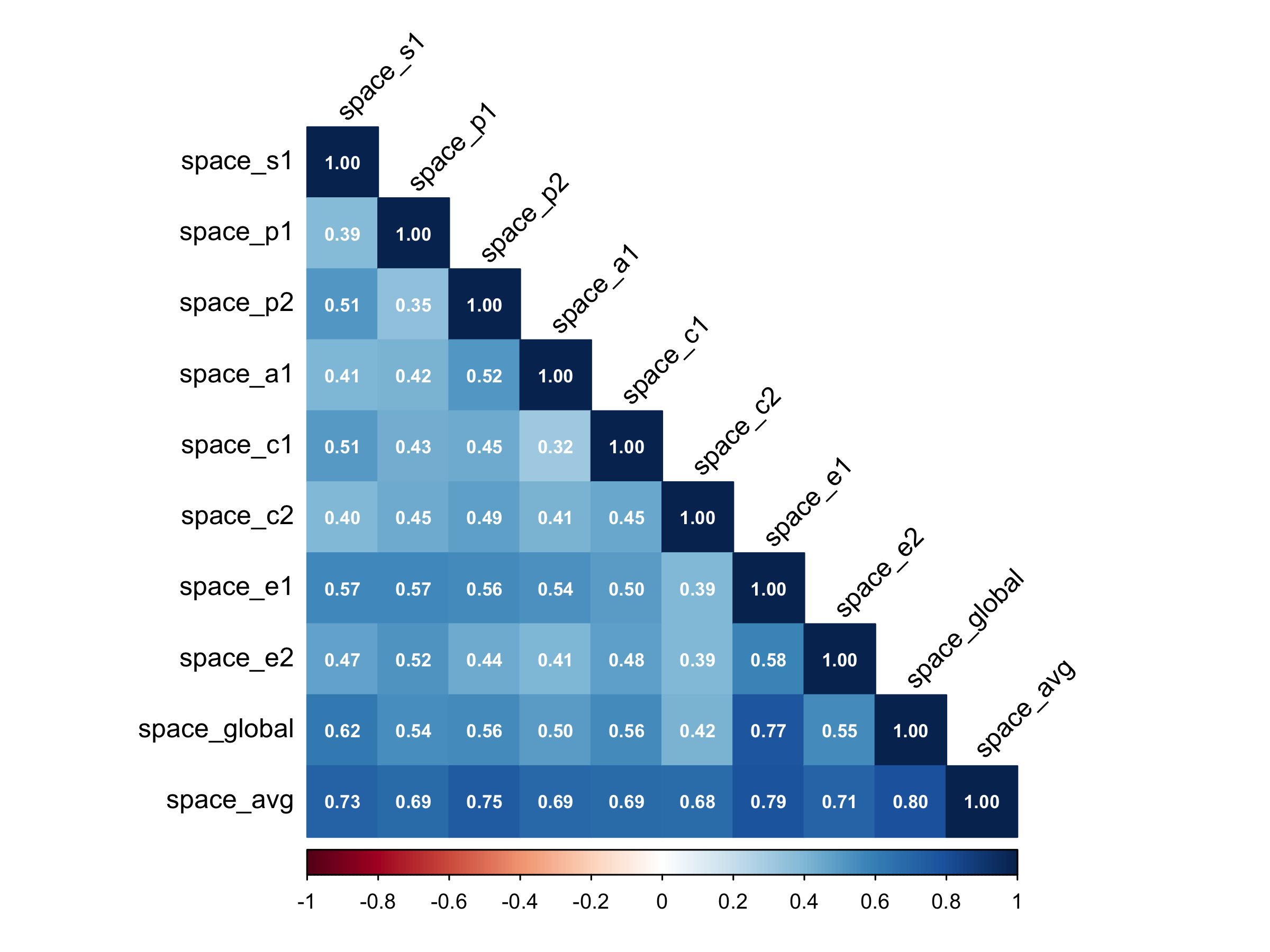}
\caption{Pairwise polychoric between items on the SPACE questionnaire. The subscripts indicate which factor in the SPACE productivity model the question addresses: satisfaction and well-being, performance, activity, communication and collaboration, efficiency and flow. Elements are listed in Table~\ref{tab:space_items}, except \texttt{space\_avg}, which is the average of the rank of ratings (on a scale of 1 to 5) recorded for each item.}\label{fig:space_correlations}
\end{figure}

\clearpage

\section{Qualitative Coding}
\begin{table}[hbt!]
\caption{Themes in ethics-related reasons for non-adoption of genAI tools}
\label{tab:ethics_subthemes}
\centering
\small
\renewcommand{\arraystretch}{1.3}
\setlength{\tabcolsep}{4pt}

\begin{tabular}{p{0.22\textwidth} p{0.28\textwidth} p{0.40\textwidth} r}
\toprule
Subtheme & Description & Example & Count\\
\midrule

Environment & Concerns about environment, climate, and sustainability related to the development or use of AI systems 
& “I will only use AI tools if I am desperate because of the high environmental cost of using AI.” 
& 25\\

General AI concerns & Ethical issues with companies and practices related to the production and use of AI tools 
& “I see serious ethical concerns in the training of these models.” 
& 24\\

Mistrust & Expresses general mistrust in AI systems 
& “Because AI is a machine learning algorithm which has serious limitations.” 
& 13\\

Degrading systems & Concern that code widely will become degraded by AI systems 
& “Increase in the production of bad code.” 
& 5\\

Professionalism & Frames AI usage as violating professional ethics 
& “I do not think that AI has any part in quality programming. I need to know what my code does, and the best way to do that is to write it myself.” 
& 5\\

Deskilling & Expresses worry that use of AI erodes programming ability broadly (beyond the individual) 
& “For scientific programming by graduate students GenAI tools is the worst thing that could ever happen...” 
& 2\\

\bottomrule
\end{tabular}
\end{table}

\end{appendices}

\clearpage

\bibliography{references}

\end{document}